\pdfoutput=1
\documentclass[10pt,journal]{IEEEtran}

\usepackage{amsmath,amsfonts}
\usepackage{algorithmic}
\usepackage{algorithm}
\usepackage{array}
\usepackage[caption=false,font=normalsize,labelfont=sf,textfont=sf]{subfig}
\usepackage{textcomp}
\usepackage{url}
\usepackage{verbatim}
\usepackage[pdftex]{graphicx}
\usepackage{cite}
\usepackage{booktabs}
\usepackage{rotating}
\usepackage{xcolor}
\usepackage{tikz}
\usetikzlibrary{arrows.meta,shapes.geometric,shapes.misc,positioning,fit,calc,backgrounds}
\usepackage{pgfplots}
\usepackage{placeins}
\pgfplotsset{compat=1.17}
\definecolor{accessblue}{cmyk}{1,0.3,0,0.2}

\begin{document}

\title{Model Compression and Hardware-Aware Acceleration for Deep Learning on FPGAs:\\A Co-Design Taxonomy and Comparative Analysis}

\author{\IEEEauthorblockN{Peter Forcha, H. Kajekusumadhar, Mbua Peter, Muhammed Kawser, Audrey Cyriell Mo,Christophe Bobda}
\IEEEauthorblockA{peter.forcha@ufl.edu, h.kajekusumadhar@ufl.edu, mbuapete@gmail.com, muhammedkawser@gmail.com, audreycyriell.mo@ufl.edu,cbobda@ece.ufl.edu
}}

\markboth
{Peter \MakeLowercase{\textit{et al.}}: Model Compression and Hardware-Aware Acceleration for Deep Learning on FPGAs}
{Peter \MakeLowercase{\textit{et al.}}: Model Compression and Hardware-Aware Acceleration for Deep Learning on FPGAs}

\maketitle

\begin{abstract}
Deploying deep neural networks on Field-Programmable Gate Arrays (FPGAs) requires joint
reasoning about model compression and hardware acceleration, however the most comprehensive
existing cross-platform treatment of this space, Deng et al.~\cite{deng2020model}, compared
compression techniques against CPU, GPU, FPGA, and ASIC targets at the level of broad,
qualitative trade-offs, and not specific FPGA resource consequences. This survey instead
restricted the scope to FPGAs alone and organized 25 compression-hardware co-design case studies
(2015--2026) into a five-category taxonomy defined by which FPGA resources each strategy
primarily reshapes: DSP-eliminating, DSP-repurposing/mixed-precision, sparsity-exploiting,
memory-hierarchy-driven, and toolchain/deployment-level. Normalizing these case studies along a
common set of dimensions (compression ratio, accuracy change, throughput, energy efficiency,
and DSP/LUT/BRAM utilization) surfaces a central, quantitative finding; of the 25 reviewed
works, only \emph{one} reported a compression ratio and accuracy change measured against a
single common baseline, and only \emph{two} reported energy efficiency normalized against a
common GPU baseline, exposing a field-wide characterization gap that no individual toolchain
(FINN, HLS4ML, Vitis AI, or DNNWeaver) resolves on its own. Building on this taxonomy and
meta-analysis, we formalize six open challenges: toolchain fragmentation,
accuracy--efficiency characterization, automated mixed-precision optimization, sparse
computation reliability, persistent memory bottlenecks, and FPGA-based training. Each is paired
with a concrete next step grounded in extending an existing, cited technique, not a
general call for future work.
\end{abstract}


\section{Introduction}
\IEEEPARstart{T}{he} exponential growth of deep learning models has revolutionized numerous
domains, including computer vision, natural language processing, and speech recognition.
Modern neural networks such as ResNet, BERT, and the GPT series have achieved unprecedented
accuracy and generalization capabilities, primarily because their massive parameter counts and
architectural depth. This trend is evident in the evolution of influential architectures: AlexNet
(2012) contain 60 million parameters~\cite{krizhevsky2012imagenet}, VGGNet (2014) expand
to 138 million parameters~\cite{simonyan2014very}, and ResNet-152 (2016) scale to 152 layers
with 60 million parameters~\cite{he2016deep}. More recently, large language models have grown
dramatically, with GPT-3 reaching 175 billion parameters~\cite{brown2020language}, and models
continue to scale into the trillion-parameter regime. Although deeper and larger networks generally
achieve higher accuracy~\cite{he2016deep,li2020train}, they also introduce significant challenges
in terms of memory usage, computational cost, and energy consumption, particularly for
real-time or embedded inference.

To address these challenges, research has increasingly focused on model compression
techniques that reduce the size, computational cost, and energy consumption of neural networks,
while maintaining accuracy. These methods are critical for enabling deep learning inference on
both edge devices, where power, latency, and resource constraints are stringent, and in data
centers, where energy efficiency and throughput are essential for a cost-effective deployment. Primary compression strategies include: \emph{pruning}, which removes redundant weights or
structures; \emph{quantization}, which reduces the numerical precision of weights and activations;
\emph{knowledge distillation}, which transfers knowledge from large teacher models to compact
student models; and \emph{low-rank factorization}, which decomposes weight matrices into
lower-dimensional representations. Additionally, weight sharing, efficient architecture design,
and neural architecture search contribute to compact model creation.

Field-Programmable Gate Arrays (FPGAs) have emerged as a promising hardware platform
for deploying compressed machine learning models, offering a reconfigurable middle ground
between the fixed-function efficiency of ASICs and the general-purpose flexibility of CPUs and
GPUs~\cite{deng2020model}; Section~\ref{sec:background} gives the architectural detail this survey
builds on. The case of FPGA acceleration is not only solely theoretical. Microsoft's Catapult deployment
reduced the Bing search latency by 25\% while roughly doubling throughput by offloading
computationally expensive ranking operations to FPGA fabric across its data
centers~\cite{caulfield2016cloudscale}. FPGA-based neural network inference accelerators have
been reported to achieve an order-of-magnitude improvement in speed and energy efficiency
over GPU baselines for state-of-the-art models~\cite{guo2019survey}. These deployed gains motivate the compression-hardware co-design problem addressed by this survey. Realizing them in
practice requires a compressed model whose sparsity pattern, numerical precision, and memory
footprint are matched to the specific FPGA resource budget available; treating compression and
hardware mapping as independent steps forfeits that match.

Despite these advantages, deploying deep neural networks directly onto FPGA hardware
remains challenging owing to limited on-chip resources, memory bandwidth constraints, and the
complexity of hardware-software co-design~\cite{guo2019survey,sze2017efficient,shantharama2020hardware}. Therefore, model compression and hardware-aware
optimization techniques therefore play a critical role in bridging the gap between complex neural
architectures and FPGA resource constraints~\cite{choudhary2020comprehensive,koilia2024hardware}.

\subsection{Scope and Novelty of the Survey}
This survey addresses the critical intersection of machine learning compression and
FPGA-based acceleration, focusing on three interrelated concepts: (1) compression techniques
(pruning, quantization, distillation and low-rank factorization); (2) hardware acceleration
strategies (dataflow architectures, systolic arrays, sparse computation, and  HLS optimizations);
and (3) AI-to-FPGA compilation flows (FINN, HLS4ML, Vitis AI, and DNNWeaver).

Deng et al.~\cite{deng2020model} provided the most comprehensive existing cross-platform
treatment of this space, surveying how compression techniques map onto CPU, GPU, FPGA, and
ASIC deployments in terms of broad trade-offs in memory bandwidth, computational throughput,
and energy efficiency. Their treatment necessarily operates at \emph{technique} granularity;
for example, quantization is discussed once, with its hardware implications summarized
qualitatively across all four platform types instead of being traced down to a specific FPGA
resource. Other surveys similarly
treat compression~\cite{cheng2017survey,choudhary2020comprehensive} and hardware acceleration
as largely separable concerns, or address them for platforms other than FPGAs.

This survey instead restricts the scope to FPGAs alone in order to go one level deeper. Instead of
asking \emph{how quantization affects hardware in general}, we ask \emph{which
specific FPGA resource (DSP, LUT, BRAM, or off-chip memory bandwidth) does a given
compression choice reshape, and through which toolchain reshaping is realized in
practice}. This FPGA-resource-level granularity is what the DSP/LUT/BRAM column in
Table~\ref{tab:master} and the five-category taxonomy of Section~\ref{sec:taxonomy} make
explicit. It is what lets this survey trace a single technique such as binarization, from an XNOR-popcount arithmetic choice (FINN) to a specific reported LUT/DSP/BRAM
utilization number, a level of detail that a cross-platform survey's necessarily coarser
treatment does not attempt.

\subsection{Survey Methodology}
The 25 primary case studies reviewed in this survey were identified through citation-chaining
from two foundational compression-hardware co-design works, Han et al.'s Deep
Compression~\cite{han2015deep} and Zhang et al.'s roofline analysis for FPGA
accelerators~\cite{zhang2015optimizing}, combined with targeted searches of FPGA-focused
venues (ACM/SIGDA FPGA, FCCM, ISCA, MICRO, ICCAD, and DATE) for works published 2015--2026
that address compression-hardware co-design specifically for FPGA deployment. This is a
\emph{targeted} literature search, not a systematic one (in the PRISMA sense), and we make
that distinction explicit here because Section~\ref{sec:metaanalysis} depends on it. A
systematic search across a fixed set of databases with pre-registered inclusion criteria would
be required before this survey's meta-analytic observations (for instance, only one of 25
reviewed works reported a compression ratio and accuracy change against a common baseline) could
be read as an estimate of the true state of the field, as opposed to an observation of the
25 studies we identified and reviewed. Inclusion required that a work (i) target FPGA deployment
specifically and not GPU, ASIC, or CPU alone, and (ii) report at least one compression
technique together with at least one hardware-level consequence of that technique. Each
included work was then evaluated uniformly across the 11 dimensions of
Table~\ref{tab:master}: publication year, taxonomy category, model/domain, hardware
platform, compression ratio, accuracy change, throughput, energy efficiency, DSP/LUT/BRAM
usage, and baseline comparison. Any dimension that the source does not report is marked explicitly
(``N/R'') instead of omitted, per the normalization discipline formalized in
Section~\ref{sec:normalization}.

\subsection{Contributions}
This survey makes three concrete contributions:
\begin{itemize}
\item \textbf{A five-category, hardware-consequence taxonomy} (Section~\ref{sec:taxonomy})
that organizes compression-hardware co-design strategies by which FPGA resource they
primarily reshaped (DSP-eliminating, DSP-repurposing/mixed-precision, sparsity-exploiting,
memory-hierarchy-driven, and toolchain/deployment-level) instead of by publication
chronology or algorithm family alone.
\item \textbf{A normalized master comparison and toolchain comparison}
(Tables~\ref{tab:master} and~\ref{tab:toolchain}) spanning 25 case studies and four deployment
toolchains (FINN, HLS4ML, Vitis AI, DNNWeaver), explicit about which metrics are and are not
reported, exposes the field's characterization gaps instead of obscuring them
(Section~\ref{sec:metaanalysis}).
\item \textbf{A concrete, six-item research agenda} (Table~\ref{tab:agenda}) in which each
open challenge is paired with a specific current state-of-the-art attempt, the stated reason it
falls short, and a next step grounded in extending an existing cited technique, not a
general call for future work.
\end{itemize}

\subsection{Paper Organization}
The remainder of this paper is organized as follows. Section~\ref{sec:background} presents the
theoretical background of machine learning compression and the FPGA architecture.
Section~\ref{sec:taxonomy} introduces the five-category taxonomy and reviews its 25 case
studies, along with Table~\ref{tab:master}. Section~\ref{sec:metaanalysis} normalizes
reported metrics across these case studies, presenting the compression-accuracy and
energy-efficiency comparisons of the data support and flagging two disagreements that surveyed
literature leaves unresolved. Section~\ref{sec:discussion} presents comparative findings on
compression techniques, hardware implications, acceleration strategy trade-offs, and
deployment toolchains (Table~\ref{tab:toolchain}). Section~\ref{sec:agenda} presents the open
challenges in the research agenda (Table~\ref{tab:agenda}). Finally Section~\ref{sec:conclusion}
concludes the paper.

\section{Background and Context}
\label{sec:background}

\subsection{Introduction to Machine Learning}

\subsubsection{Machine Learning Overview}
Machine Learning (ML) is a branch of Artificial Intelligence (AI) that enables computer
systems to learn patterns from data without  explicit programming. ML algorithms
analyze both structured and unstructured data to identify patterns, make predictions, and
support informed decision-making.

\subsubsection{Categories of Machine Learning}
ML is broadly categorized into three main classes:
\begin{itemize}
  \item \textbf{Supervised Learning:} Models trained on labeled data for regression
  (continuous prediction) or classification (categorical outcomes)~\cite{goodfellow2016deep}.
  \item \textbf{Unsupervised Learning:} Models learn from unlabeled data for clustering,
  dimensionality reduction, and anomaly detection~\cite{goodfellow2016deep}.
  \item \textbf{Reinforcement Learning:} An agent interacting with an environment,
  receiving rewards or penalties, to maximize cumulative rewards~\cite{sutton2018reinforcement}.
\end{itemize}
Additional paradigms include semi-supervised learning (pseudo-labeling) and
self-supervised learning, in which the system generates its own supervisory signals (e.g.,
SimCLR, BERT, and MAE).

\subsubsection{Deep Learning}
\begin{figure*}[tp]
  \centering
  \includegraphics[width=\textwidth]{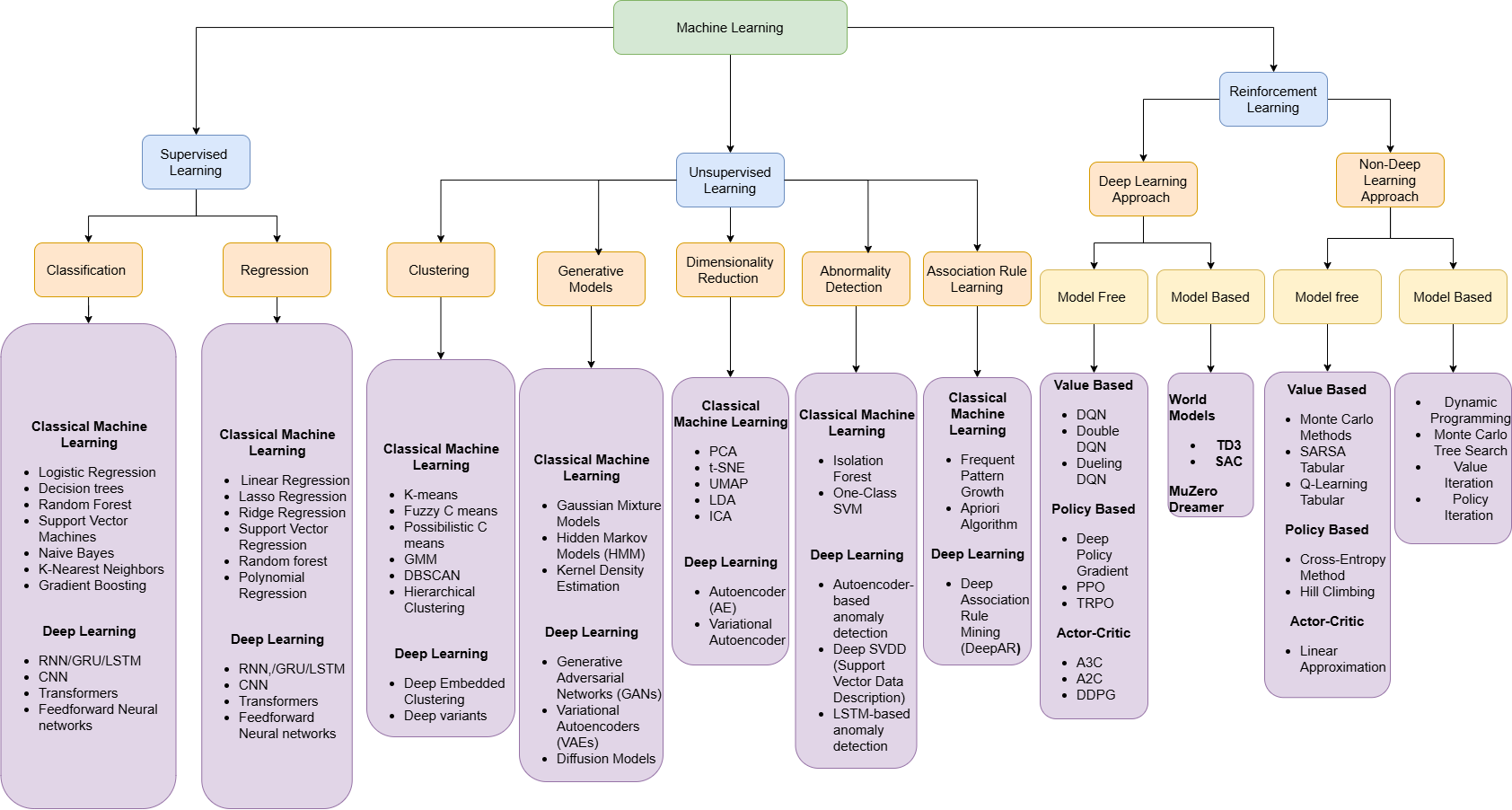}
  \caption{Complete Machine Learning Taxonomy: Deep Learning and Traditional Approaches.}
  \label{fig:ml_taxonomy}
\end{figure*}

Deep Learning (DL) is a specialized ML subfield that employ artificial neural networks (ANNs)
inspired by the human brain. Stacking many layers of perceptrons forms deep neural
networks (DNNs) that are capable of learning hierarchical representations.
Fig.~\ref{fig:ml_taxonomy} illustrates the ML taxonomy and deep learning techniques
applicable to each class.

\subsubsection{Machine Learning Workflow}
\begin{figure}[tbp]
  \centering
  \includegraphics[width=0.82\columnwidth]{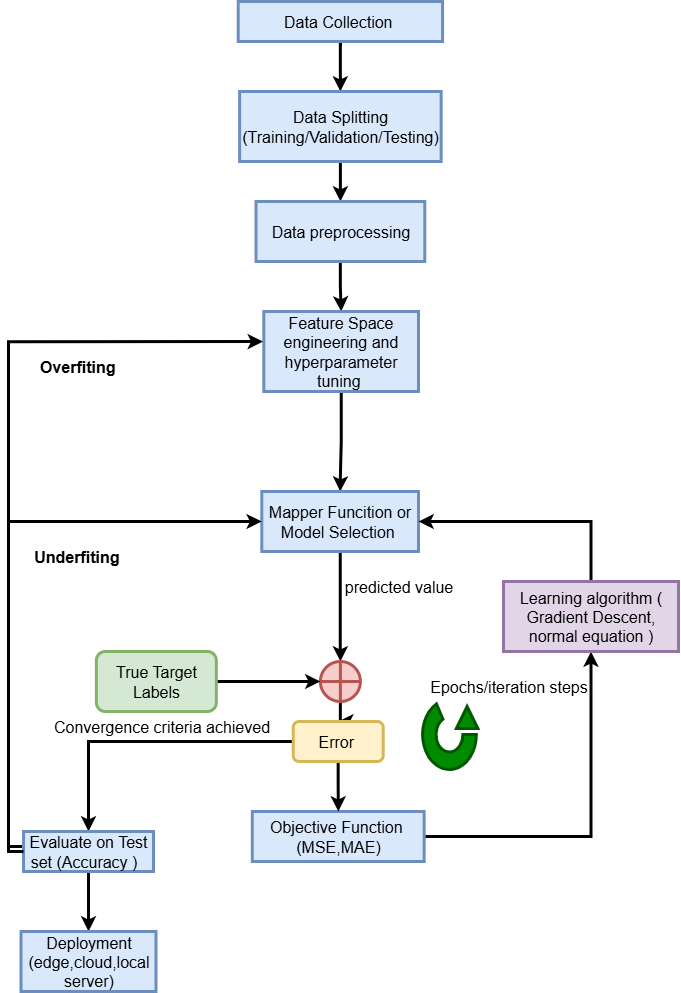}
  \caption{A typical machine learning workflow showing key stages from data collection to
  model deployment.}
  \label{fig:ml_workflow}
\end{figure}

A typical ML workflow (Fig.~\ref{fig:ml_workflow}) begins with \emph{data collection},
which forms the foundation for training and evaluation~\cite{goodfellow2016deep}. Feature
engineering constructs at input representation (e.g., polynomial expansion and RBF
mapping)~\cite{kuhn2013applied}. Preprocessing handles missing values, outliers, and feature
scaling (standardization/normalization) to improve the convergence~\cite{han2022data}. The
dataset was partitioned into \emph{training}, \emph{validation}, and \emph{test} subsets,
with all transformations derived exclusively from the training data to prevent leakage~\cite{lecun2015deep}.

During \emph{model training}, a loss function quantifies the prediction error, and optimizers such
as SGD, Adam, and RMSProp minimize it~\cite{goodfellow2016deep}. The \emph{parameters}
(weights, biases) are learned from the data, whereas \emph{hyperparameters} (learning rate, batch
size, depth) are set before training. Cross-validation guides hyperparameter
selection~\cite{bergstra2012random} and mitigates overfitting (high training, poor test
performance) and underfitting (models that are too simple to capture trends). As models scale to
billions of parameters, compression and hardware-aware optimization have become essential for
resource-constrained deployment~\cite{cheng2017survey,deng2020model}.

\subsubsection{Neural Network Basics}
\begin{figure}[tbp]
  \centering
  \includegraphics[width=\columnwidth]{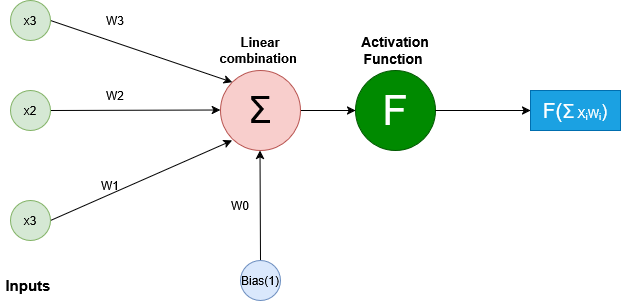}
  \caption{Single artificial neuron: weighted inputs summed with bias, passed through an
  activation function to produce output $\hat{y}$.}
  \label{fig:neuron}
\end{figure}

\begin{figure}[tbp]
  \centering
  \includegraphics[width=\columnwidth]{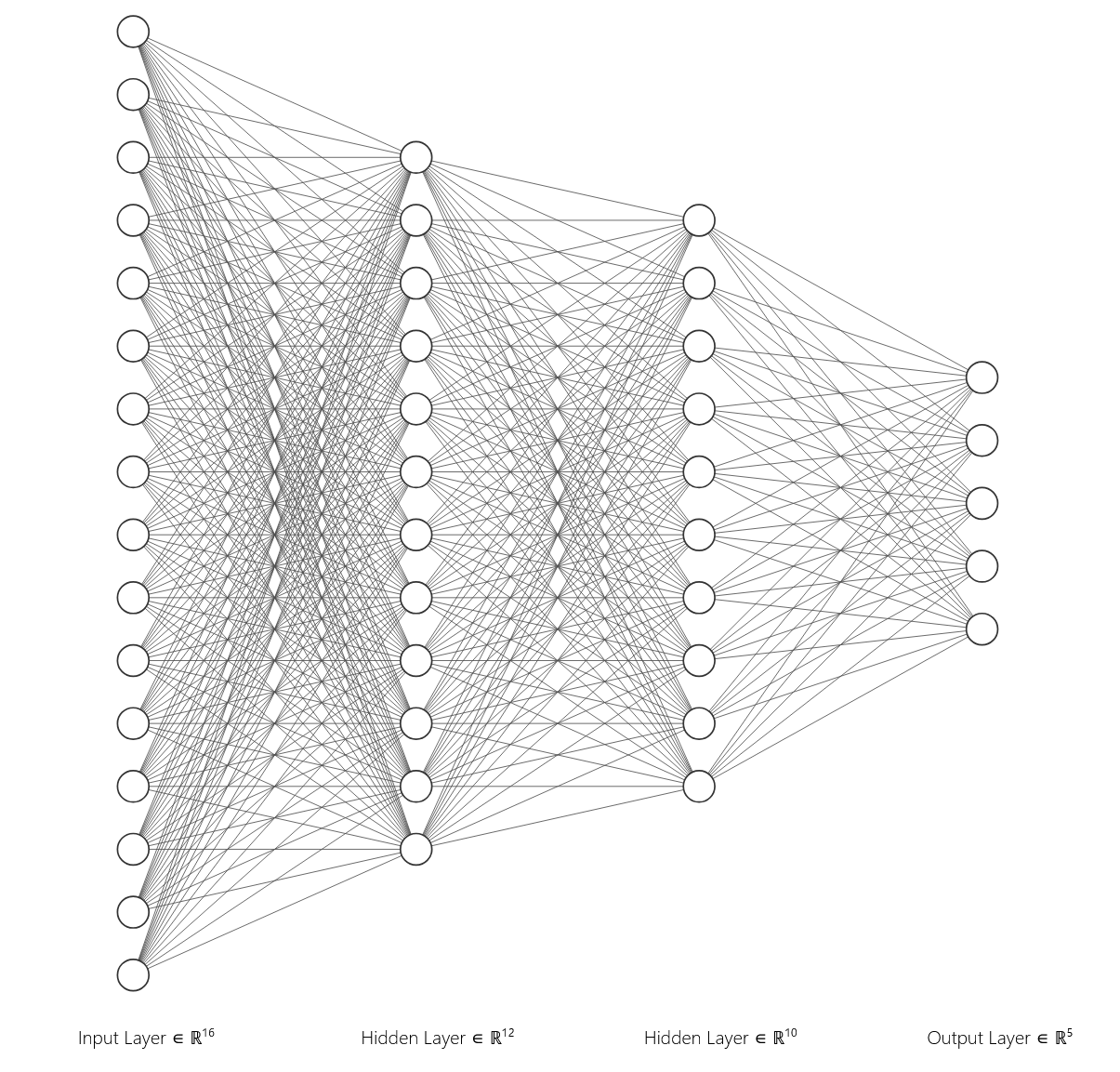}
  \caption{Fully connected feedforward neural network with two hidden layers.}
  \label{fig:fcnn}
\end{figure}

The perceptron is the fundamental building block of a DNN, which computes the weighted sum of
inputs plus bias and passes the result through an activation function
(Fig.~\ref{fig:neuron})~\cite{lecun1989backpropagation}. Multiple connected perceptrons
form a \emph{Multilayer Perceptron} (MLP)~\cite{goodfellow2016deep}. Weights are learned
via SGD, Adam, or RMSProp, and the layers between the input and output are \emph{hidden layers}. When
every neuron in one layer is connected to every neuron in the next layer, the network is \emph{fully
connected} or \emph{dense} (Fig.~\ref{fig:fcnn}); \emph{sparse} networks limit these
connections via design or pruning. Preventing overfitting requires regularization (weight
decay and  dropout) and hyperparameter tuning~\cite{goodfellow2016deep}.

Stacking these neurons into layers makes the fully connected topology of
Fig.~\ref{fig:fcnn} the dominant source of parameter count, and therefore the dominant target of
compression, in classical DNNs: a single dense layer connecting $m$ inputs to $n$ outputs
contributes $mn$ weights, so the parameter count grows quadratically with the layer width even before
depth is considered~\cite{goodfellow2016deep}. This is precisely the structure that pruning and
low-rank factorization, discussed later in this section, target: removing or approximating
individual connections in a fully connected layer maps directly onto a reduction in the $mn$
weight matrix, with no change required for the input/output behavior of the network.

\subsubsection{Common Architectures}
DNNs have evolved through several key paradigms:
\begin{itemize}
  \item \textbf{Feedforward NNs:} The foundational multilayer perceptron established
  gradient-based learning~\cite{lecun1989backpropagation}, with information flowing
  unidirectionally from the input to the output through successive layers.
  \item \textbf{CNNs:} Exploit spatial structure via local connectivity, weight sharing, and
  pooling~\cite{lecun1998gradient}, with LeNet, AlexNet~\cite{krizhevsky2012imagenet},
  VGGNet~\cite{simonyan2014very}, and ResNet~\cite{he2016deep} to achieve human-level
  performance in image classification.
  \item \textbf{RNNs/LSTMs:} Handle sequential data. LSTM~\cite{hochreiter1997long} and
  GRU~\cite{cho2014learning} use gating mechanisms to overcome vanishing gradients.
  \item \textbf{Transformers:} Replace recurrence with self-attention~\cite{vaswani2017attention},
  build on earlier attention ~\cite{bahdanau2014attention}, and underpin
  BERT~\cite{devlin2019bert}, GPT-2/3~\cite{radford2019language,brown2020language},
  LLaMA~\cite{touvron2023llama}, and Vision Transformers~\cite{dosovitskiy2021image}.
  \item \textbf{Generative Models:} GANs~\cite{goodfellow2014generative},
  VAEs~\cite{kingma2014auto}, and diffusion models~\cite{ho2020denoising,song2021denoising}
  have achieved state-of-the-art data syntheses.
  \item \textbf{GNNs:} Graph Neural Networks~\cite{scarselli2008graph,kipf2017semi} extend
  deep learning to graph-structured data, with Graph Attention Networks~\cite{velickovic2018graph}
  and GraphSAGE~\cite{hamilton2017inductive}, finding broad applications.
\end{itemize}
Understanding these architectures is essential for compression and FPGA deployment, as each
presents unique optimization opportunities and constraints~\cite{sze2017efficient}.

\subsubsection{Computational and Memory Requirements}
The DNN resource demands depend on the architectural complexity, task difficulty, and operations
employed~\cite{canziani2017analysis}. The type of operation matters: matrix multiplications
in fully-connected layers have different memory access patterns from convolutions, whereas
activations, pooling, and batch normalization each add unique computational and memory
overhead~\cite{canziani2017analysis,culurciello2018computation}. Memory demand is
critically phase-dependent; \emph{training} retains all forward-pass activations for
backpropagation (e.g., ResNet-50 produces $\sim$16 million activations per pass, all stored
during training)~\cite{chen2016sublinear}, whereas \emph{inference} requires only model
weights and temporary single-pass storage, typically $3$--$5\times$ less memory. For FPGAs,
inference-only deployment enables the acceleration of models that are infeasible to train on-devices.

\subsection{Machine Learning Model Compression Techniques}
\begin{figure*}[tp]
  \centering
  \resizebox{\textwidth}{!}{%
  \begin{tikzpicture}[
    root/.style={draw, fill=accessblue!15, font=\small\bfseries, align=center, minimum height=0.6cm},
    cat/.style={draw, fill=gray!10, font=\scriptsize\bfseries, align=center, minimum height=0.5cm, text width=2.6cm},
    leaf/.style={draw, fill=blue!5, font=\scriptsize, align=center, minimum height=0.42cm, text width=2.6cm},
    >=stealth
  ]
    \node[root] (root) at (8,5.6) {Model Compression};
    \node[cat] (quant) at (0,4.0)  {Quantization};
    \node[cat] (prune) at (3.2,4.0) {Pruning};
    \node[cat] (lrf)   at (6.4,4.0) {Low-Rank\\Factorization};
    \node[cat] (kd)    at (9.6,4.0) {Knowledge\\Distillation};
    \node[cat] (ws)    at (12.8,4.0) {Weight Sharing\\\& Clustering};
    \node[cat] (hyb)   at (16.0,4.0) {Hybrid\\Approaches};
    \foreach \c in {quant,prune,lrf,kd,ws,hyb} { \draw[->] (root) -- (\c); }

    \node[leaf] (q1) at (0,3.0) {Granularity: layer-, channel-, or group-wise};
    \node[leaf] (q2) at (0,2.0) {Precision level: INT4, INT8, or binary};
    \node[leaf] (q3) at (0,1.0) {Training: QAT vs.\ PTQ};
    \node[leaf] (q4) at (0,0.0) {Scope: weight, activation, or both};
    \node[leaf] (q5) at (0,-1.0) {Zero point: symmetric vs.\ asymmetric};
    \draw[->] (quant) -- (q1); \draw[->] (q1) -- (q2); \draw[->] (q2) -- (q3);
    \draw[->] (q3) -- (q4); \draw[->] (q4) -- (q5);

    \node[leaf] (p1) at (3.2,3.0) {Granularity: structured vs.\ unstructured};
    \node[leaf] (p2) at (3.2,2.0) {Selection: magnitude-, gradient-, or entropy-based};
    \draw[->] (prune) -- (p1); \draw[->] (p1) -- (p2);

    \node[leaf] (l1) at (6.4,3.0) {$\mathbf{W}\!\approx\!\mathbf{U}\mathbf{V}^{\top}$, rank $r$ selection};
    \draw[->] (lrf) -- (l1);

    \node[leaf] (k1) at (9.6,3.0) {Teacher--student: feature-, logit-, or relation-based; ensemble/multi-teacher};
    \node[leaf] (k2) at (9.6,2.0) {Self-distillation: online or snapshot (no external teacher)};
    \draw[->] (kd) -- (k1); \draw[->] (k1) -- (k2);

    \node[leaf] (w1) at (12.8,3.0) {Clustering algorithm: $k$-means or GMM; hashed weight sharing};
    \draw[->] (ws) -- (w1);

    \node[leaf] (h1) at (16.0,3.0) {Combines techniques: quantization+pruning; distillation+quantization};
    \draw[->] (hyb) -- (h1);
  \end{tikzpicture}%
  }
  \caption{Taxonomy of model compression techniques covered in this survey. The five core
  techniques (quantization, pruning, low-rank factorization, knowledge distillation, and
  weight sharing) are each developed in their own subsection below; hybrid approaches combine
  two or more of these directly. This taxonomy culminates in the hardware-consequence
  taxonomy of Section~\ref{sec:taxonomy} that reorganizes these same techniques by which FPGA
  resource they primarily reshape.}
  \label{fig:compression_taxonomy}
\end{figure*}
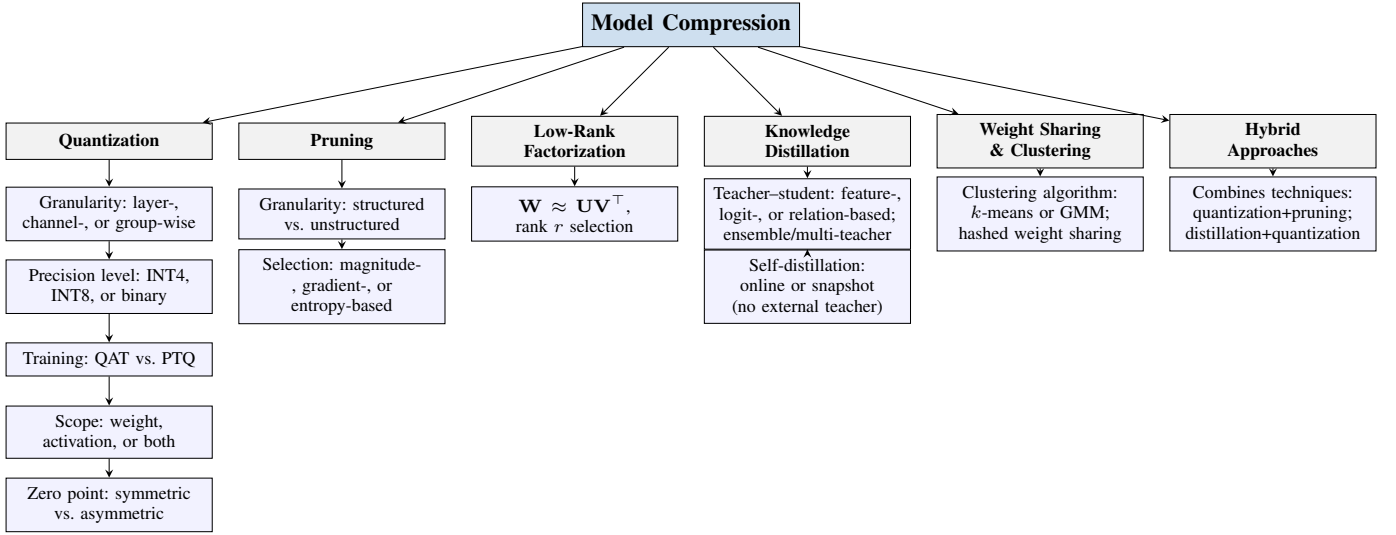

Model compression reduces the size and computational complexity of the model while maintaining
accuracy, enabling efficient deployment on resource-constrained platforms.
Fig.~\ref{fig:compression_taxonomy} shows the taxonomy of the primary compression methods.

\subsubsection{Quantization}
Quantization maps high-precision values to lower-precision discrete
representations~\cite{gholami2022survey}, thereby reducing the bits needed for weights and/or
activations. Reducing precision lowers the memory footprint but can degrade
accuracy~\cite{hashemi2017understanding}. Key dimensions:
\begin{itemize}
  \item \textbf{Precision level:} FP16 down to INT4 or binary; mixed-precision schemes
  combining multiple formats~\cite{gholami2022survey}.
  \item \textbf{Timing:} \emph{Quantization-Aware Training} (QAT) integrates quantization
  during training, adapting to precision loss; \emph{Post-Training Quantization} (PTQ) converts
  after training, a simpler route that typically incurs higher accuracy
  degradation~\cite{frantar2022gptq}.
  \item \textbf{Granularity:} Layer-wise, channel-wise, or group-wise quantization parameters.
  \item \textbf{Symmetry:} Symmetric vs.\ asymmetric zero-point; uniform vs.\ non-uniform
  bin spacing.
\end{itemize}

\subsubsection{Pruning}
Pruning eliminates weights, neurons, or structural components that contribute
minimally to the accuracy~\cite{han2015learning}. Two main categories:
\begin{itemize}
  \item \textbf{Unstructured pruning}~\cite{dong2017learning,hassibi1992second,xiao2019autoprune}:
  removes individual weights and achieve high compression ratios but yields irregular sparse
  matrices that are difficult to exploit on regular hardware such as systolic arrays.
  \item \textbf{Structured pruning}~\cite{he2018amc,huang2018data,lin2018accelerating,yu2022hessian}:
  eliminates all neurons, channels, or layers, preserving regular computation patterns
  conducive to an efficient FPGA implementation.
\end{itemize}
These criteria include magnitude-based thresholds~\cite{han2015learning},
entropy-based measures~\cite{hur2019entropy}, and voting-based
approaches~\cite{alqahtani2021neuron}.

\subsubsection{Low-Rank Factorization}
Low-Rank Factorization (LRF) approximates a weight matrix
$\mathbf{W}\!\in\!\mathbb{R}^{m\times n}$ as
$\mathbf{W}\!\approx\!\mathbf{U}\mathbf{V}^{\top}$, reducing parameters from $mn$ to
$r(m+n)$ where $r\!\ll\!\min(m,n)$~\cite{ren2024low}. Despite its theoretical advantages, LRF
is less widely adopted: it alters network architecture, disrupts optimized dense matrix
routines, and often requires custom kernels to realize inference speedup~\cite{yi2021towards}.
The rank selection problem, which determines the optimal $r$ per layer, further increases deployment
complexity.

\subsubsection{Knowledge Distillation}
\begin{figure}[tbp]
  \centering
  \includegraphics[width=0.9\columnwidth]{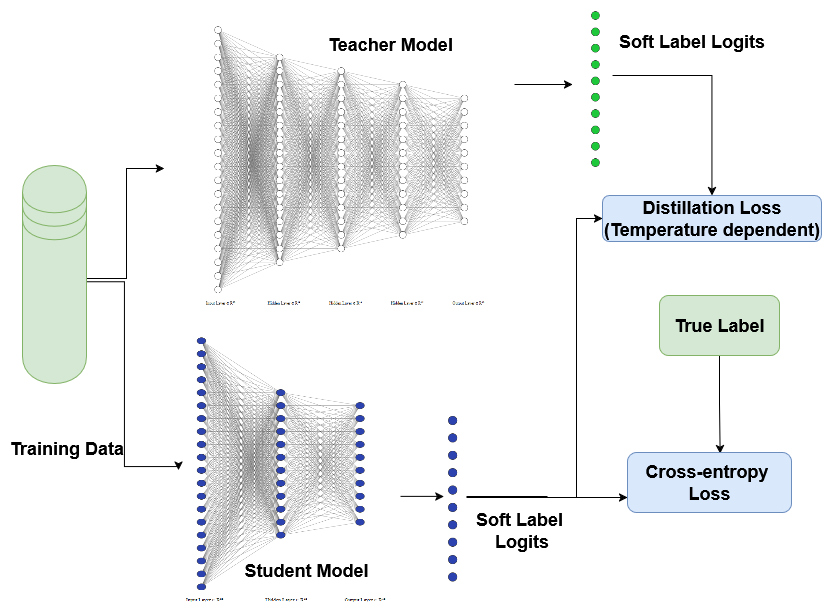}
  \caption{Classical Knowledge Distillation: a student network is trained to mimic
  softened outputs from a larger teacher network.}
  \label{fig:kd_diagram}
\end{figure}

Knowledge Distillation (KD) trains a smaller \emph{student} network to mimic the softened
output distribution of a larger pretrained \emph{teacher}~\cite{hinton2015distilling,gou2021knowledge,li2017learning,mishra2017apprentice}
(Fig.~\ref{fig:kd_diagram}). The standard KD loss minimizes the KL divergence between the
temperature rescaled softmax outputs:
\begin{equation}
\mathcal{L}_{\text{KD}}^{\text{KL}} = T^2 \sum_{c} \sigma\!\left(\tfrac{z^t}{T}\right)_c
\log \frac{\sigma\!\left(\tfrac{z^t}{T}\right)_c}{\sigma\!\left(\tfrac{z^s}{T}\right)_c}
\end{equation}
Alternatively, MSE between softened logits:
\begin{equation}
\mathcal{L}_{\text{KD}}^{\text{MSE}} = \left\|\sigma\!\left(\tfrac{z^t}{T}\right) -
\sigma\!\left(\tfrac{z^s}{T}\right)\right\|_2^2
\end{equation}
The total objective blends distillation with cross-entropy:
\begin{equation}
\mathcal{L}_{\text{total}} = \alpha\cdot\mathcal{L}_{\text{KD}} + (1-\alpha)\cdot\mathcal{L}_{\text{CE}}
\end{equation}
Extensions include Relational KD~\cite{park2019relational} (pairwise embedding structure),
feature-level alignment~\cite{chen2020learning}, self-distillation without an external
teacher~\cite{yang2023knowledge}, and theoretical analysis of why soft labels improve over
hard ones~\cite{mandal2024theoretical}.

\FloatBarrier

\subsubsection{Weight sharing and clustering}
Weight sharing reduces the memory footprint by clustering similar weights and replacing them
with shared centroids. Han et al.~\cite{han2015deep} applied $k$-means clustering, and
Chen et al.~\cite{chen2015compressing} employed hash functions to implicitly enforce sharing
without explicit clustering.

\subsection{Introduction to FPGA}
Field-Programmable Gate Arrays (FPGAs) are semiconductor devices that can be reconfigured after
manufacturing to perform a wide range of computational tasks. Unlike fixed-function ASICs
or general-purpose CPUs/GPUs, an FPGA comprises programmable logic blocks, configurable
interconnects, and I/O blocks that implement an arbitrary digital logic. Specifically, an FPGA is
built from Look-Up Tables (LUTs) that realize combinational logic, Digital Signal Processing (DSP)
units that realize fixed-point and floating-point arithmetic, on-chip memory (BRAM, or HBM on
newer data-center parts) that holds weights and activations close to the compute fabric, and a
programmable interconnect that wires these blocks together. Unlike a CPU or ASIC, this fabric can
be reprogrammed post-manufacturing to match the dataflow of a specific compressed neural
network, which allows compression choices to translate directly into resource savings, rather
than remaining a purely algorithmic exercise~\cite{deng2020model}.

\subsubsection{FPGA Architecture and Components}
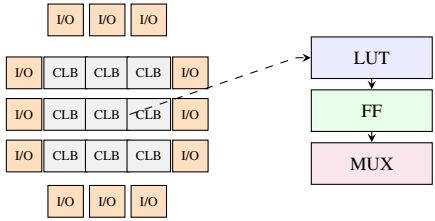
\begin{figure}[tbp]
  \centering
  \begin{tikzpicture}[
    io/.style={draw, fill=orange!25, minimum width=0.42cm, minimum height=0.42cm, font=\tiny},
    clb/.style={draw, fill=gray!12, minimum width=0.42cm, minimum height=0.42cm, font=\tiny},
    block/.style={draw, minimum height=0.55cm, font=\scriptsize, align=center},
    >=stealth
  ]
    \foreach \x in {0,1,2} {
      \node[io] (iot\x) at (\x*0.55+0.55,1.8) {I/O};
      \node[io] (iob\x) at (\x*0.55+0.55,-0.6) {I/O};
    }
    \foreach \y in {0,1,2} {
      \node[io] (iol\y) at (0,\y*0.55+0) {I/O};
      \node[io] (ior\y) at (2.2,\y*0.55+0) {I/O};
    }
    \foreach \x in {0,1,2} {
      \foreach \y in {0,1,2} {
        \node[clb] (c\x\y) at (\x*0.55+0.55,\y*0.55) {CLB};
      }
    }
    \node[block, fill=blue!8, minimum width=1.6cm] (lut) at (4.6,1.3) {LUT};
    \node[block, fill=green!10, minimum width=1.6cm] (ff) at (4.6,0.6) {FF};
    \node[block, fill=purple!10, minimum width=1.6cm] (mux) at (4.6,-0.1) {MUX};
    \draw[->] (lut) -- (ff);
    \draw[->] (ff) -- (mux);
    \draw[->, dashed] (c11.east) -- (3.6,1.3) -- (lut.west);
  \end{tikzpicture}
  \caption{General FPGA floorplan: an array of Configurable Logic Blocks (CLBs) surrounded by
  a ring of I/O blocks and connected through a programmable interconnect; each CLB internally
  chains a Look-Up Table (LUT), a flip-flop (FF), and an output multiplexer (MUX). Original
  diagram, following the standard CLB/switch-box/I/O floorplan convention used across FPGA
  architecture literature~\cite{ahmed2022multitenant}.}
  \label{fig:fpga_arch}
\end{figure}

Modern FPGAs integrate configurable logic blocks (CLBs), programmable interconnect
networks, embedded memory blocks (BRAM), digital signal processing (DSP) units, and I/O
blocks. Fig.~\ref{fig:fpga_arch} shows the general architecture.

\begin{figure}[tbp]
  \centering
  \includegraphics[width=0.85\columnwidth]{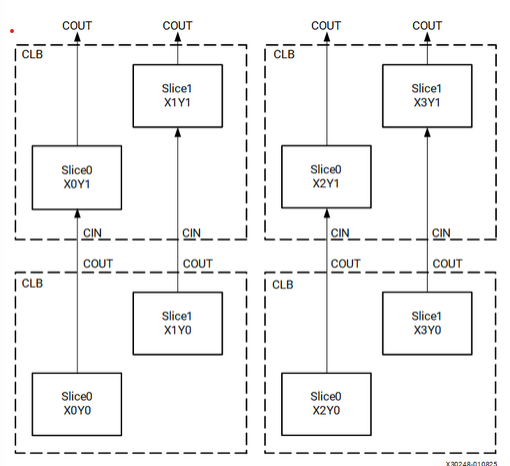}
  \caption{AMD (Xilinx) 7 Series CLB architecture: two slices each with four 6-input LUTs
  and eight flip-flops per slice. Image source: AMD UG474~\cite{amd2016clb}.}
  \label{fig:amd_clb}
\end{figure}

\paragraph{Modern FPGA Logic-Block Architecture}
Regardless of the vendor, the fundamental computational unit of a modern FPGA is a small
lookup-table-plus-register block replicated across the fabric and stitched together by dedicated
carry chains for fast arithmetic. AMD's Configurable Logic Block (CLB), for instance, packs two
slices per CLB, each with four 6-input LUTs and eight flip-flops, with SLICEM slices additionally
usable as distributed memory and dedicated carry chains avoiding general routing for arithmetic
paths~\cite{amd2016clb} (Fig.~\ref{fig:amd_clb}); Intel's Adaptive Logic Module (ALM) reaches a
comparable density through a different decomposition, fracturing a single 8-input LUT into two
independent 4-input functions (or other combinations) alongside two dedicated full adders and
four registers, with combinational, arithmetic, and shared-arithmetic operating
modes~\cite{intel2024alm}. The two vendors' particular slice/ALM bit-counts matter less for this
survey than  they have in common: on-chip memory (Block RAM, on-chip data and lookup-table
storage), DSP blocks (dedicated multiply-accumulate units), a multi-level interconnect linking
logic, memory, and DSPs, clock management circuitry (PLLs and clock networks), and configurable
I/O~\cite{amd2016clb,intel2024alm}. It is this common resource inventory (LUTs, flip-flops,
DSPs, and BRAM) that the compression-hardware co-design taxonomy of
Section~\ref{sec:taxonomy} is organized around, independent of which vendor's logic block is 
realized.

\subsubsection{Reconfigurability and Parallelism}

\paragraph{Reconfigurability}
Unlike an ASIC, whose logic is fixed at fabrication, an FPGA can be reprogrammed after
manufacturing and, in modern devices, even while the rest of the chip continues to
operate~\cite{guo2019survey}. A \emph{full reconfiguration} replaces the entire design,
whereas a \emph{partial reconfiguration} swaps only a designated region of the fabric while
the remainder of the circuit keeps running~\cite{ahmed2022multitenant}, a capability that
cloud FPGA deployments in particular rely on to time-share a device across tenants or
workloads. Realizing a design on this reconfigurable fabric follows a fixed toolchain: HDL
(VHDL/Verilog) is simulated, synthesized into a technology-specific netlist, and then placed
and routed into a bitstream that configures the physical logic and interconnect. Because this
whole cycle happens without touching hardware, FPGAs support field updates, hardware reuse
across projects, and rapid design iteration in a way that fixed-function silicon cannot.

\paragraph{Parallelism}
The same reconfigurability that allows an FPGA's logic to be redrawn also allows its parallelism to be
shaped to the workload, rather than being fixed by an instruction-set pipeline as on a CPU or
GPU~\cite{sze2017efficient}. This shows up at several granularities simultaneously:
\emph{fine-grained} bit-level parallelism, where an operation such as a 16-bit addition is
realized directly in parallel hardware instead of being serialized through an ALU;
\emph{coarse-grained} parallelism, where many processing elements are replicated across the
fabric to operate on different data simultaneously, as in the systolic-array designs used by
early parallel accelerators~\cite{kung1982systolic}; and \emph{pipeline} parallelism, where a
deep datapath is broken into stages so that a new result is produced every clock cycle once the
pipeline is full. In practice, the degree of parallelism that a design can extract is bounded by how
much of the chip's LUT, DSP, and BRAM budget it is willing to spend on replication.

\subsubsection{FPGA Advantages of ML Workloads}
This combination of reconfigurability and workload-shaped parallelism makes FPGAs
attractive for ML inference. Because the datapath itself is redrawn per design,
it can be built at exactly the numerical precision a compressed model needs rather than the
fixed 32-bit or 8-bit datapaths a CPU or GPU offers, so a quantized or binarized network spends
silicon only on the bits it actually uses; the FINN and Bit Fusion accelerators demonstrate
this directly, reporting substantial throughput and energy gains from custom low-precision and
mixed-precision datapaths that a general-purpose processor cannot express in
hardware~\cite{umuroglu2017finn,sharma2018bit}. That same precision-matching lowers energy
consumption, because operations are executed only at the bit-width the accuracy target requires
rather than the wider width a general-purpose ALU would spend regardless of
need~\cite{sze2017efficient}. The pipeline and dataflow architectures surveyed in
Section~\ref{sec:taxonomy}, such as fpgaConvNet~\cite{venieris2018fpgaconvnet} and
DNNBuilder~\cite{zhang2018dnnbuilder}, further show that deeply pipelined FPGA datapaths
achieve low and predictable per-inference latency, which is what real-time deployment
requires and GPU batch scheduling does not guarantee. FPGA on-chip memory can likewise be
organized into buffers tailored to a specific layer's reuse pattern instead of a fixed cache
hierarchy. The roofline-guided loop-tiling analysis by Zhang et al.~\cite{zhang2015optimizing}
and the dataflow-driven design of Eyeriss~\cite{chen2016eyeriss} both show that this custom
buffering directly translates into reduced off-chip DRAM traffic, which is typically the
dominant energy cost in DNN inference. Finally, because the same fabric can be re-targeted
by design, FPGA accelerators are model-adaptive: toolchains such as
AutoSA~\cite{wang2021autosa} and HeteroCL~\cite{lai2019heterocl} allows designers to tune dataflow
scheduling, loop tiling, and operator fusion per model rather than being locked into one
fixed architecture.

\subsection{FPGA Acceleration Techniques}
Hardware acceleration offloads heavy computations from general-purpose CPUs to specialized
hardware. FPGAs, along with GPUs and TPUs, excel at data-parallel tasks and offer
reconfigurable datapaths that can be co-optimized with ML compression techniques.

\subsubsection{Pipelining and Parallel Processing}
FPGAs implement deeply pipelined datapaths while simultaneously executing multiple operations
across stages. Zhang et al.~\cite{zhang2015optimizing} demonstrated that loop pipelining
combined with loop unrolling can achieve up to 61.62\,GFLOPS Xilinx Virtex-7 CNNs.
Angel-Eye~\cite{guo2017angel} exploited task and data-level parallelism through
careful pipeline design and parallel PE allocation. Minimizing pipeline stalls and using double
buffering to hide memory latency are key to achieving a near-optimal pipeline efficiency.

\subsubsection{Systolic Arrays and Matrix Multiplication}
Systolic arrays consist of a grid of PEs that rhythmically compute and pass data to neighbors,
enabling a high computational density and data reuse~\cite{kung1982systolic}. Google
TPU's $256\!\times\!256$ systolic array~\cite{jouppi2017datacenter} demonstrated the
approach at scale, keeping weights stationary and streaming activation. fpgaConvNet~\cite{venieris2018fpgaconvnet}
adapted this to FPGAs with automated framework support, dynamically sizing arrays per layer
to improve the efficiency. Systolic regularity also simplifies the integration of quantization, beacuse all
PEs can be uniformly modified for reduced-precision arithmetic.

\subsubsection{Memory Hierarchy Optimization}
The memory bandwidth is a critical bottleneck in FPGA-based accelerators. Ma et
al.~\cite{ma2017optimizing} showed that intelligent tiling strategies reduce the DRAM bandwidth
by up to $10\times$. Zhang et al.~\cite{zhang2018dnnbuilder} introduced DNNBuilder, achieving
$2.5\times$ energy efficiency improvement through specialized layer-specific buffering.
Efficient design requires understanding per-layer data access patterns and exploiting the temporal
and spatial localities. Emerging non-volatile memories (MRAM, ReRAM) offer additional
opportunities for hybrid memory hierarchies that balance density, latency, and power.

\subsubsection{Low-Precision Arithmetic and DSP Utilization}
FPGAs implement arbitrary-precision arithmetic, which is  unavailable in fixed-function accelerators.
FINN~\cite{umuroglu2017finn} deploys 1-bit binary networks using XNOR-based LUT
arithmetic, eliminating DSP blocks entirely and achieving up to 12,190\,fps on ImageNet
on a Xilinx PYNQ-Z1. Modern DSP blocks (e.g., Xilinx DSP48E2) support multiple precision
modes. Bit Fusion~\cite{sharma2018bit} dynamically composes narrow-precision operations
within wider DSP blocks: a single 27$\times$18-bit DSP performs two 9$\times$18-bit or four
4$\times$18-bit multiplications in parallel, enabling per-layer or per-channel precision
adaptation.

\subsubsection{Dataflow Optimization and Task-Level Parallelism}
Dataflow architectures map computational graphs directly to hardware pipelines, executing
operations as soon as inputs arrive instead of following a sequential instruction fetch. Guo et
al.~\cite{guo2019survey} showed that FPGA dataflow implementations achieve 5--10$\times$
better energy efficiency than GPUs for inference by eliminating instruction overhead and
enabling fine-grained pipelining. Operator fusion (convolution + batch normalization +
activation) reduces intermediate data movement and improves throughput by up to
40\%~\cite{qi2018paleo}, and is a key optimization in frameworks such as FINN and Vitis AI.

\section{A Taxonomy of Compression-Hardware Co-Design Strategies}
\label{sec:taxonomy}

\subsection{Related Surveys}

\subsubsection{Model Compression Surveys}
Han et al.~\cite{han2015deep} established at foundation for integrated compression by combining
pruning, trained quantization, and Huffman coding into a three-stage pipeline. Liu et
al.~\cite{liu2025survey} and Dantas et al.~\cite{dantas2024comprehensive} provided chronological
overviews and structured taxonomies examining compression-accuracy-cost trade-offs. Domain-specific
surveys address NLP models~\cite{gupta2022compression}, large language
models~\cite{zhu2024survey}, and transformers~\cite{tang2024survey}.

\subsubsection{Hardware Acceleration Surveys}
Deng et al.~\cite{deng2020model} surveyed the compression-to-hardware deployment across CPUs,
GPUs, FPGAs, and ASICs. Silvano et al.~\cite{silvano2025survey} examined deep learning
accelerators for heterogeneous HPC platforms. Koilia and Kachris~\cite{koilia2024hardware}
specifically compared FPGA, GPU, and ASIC implementations for large language model inference.

\subsubsection{Gaps in the existing literature}
Existing surveys treat compression and acceleration as separate optimization problems, providing
limited attention to FPGA-specific co-design and underexploring the translation from compressed
models to synthesizable implementations via frameworks such as FINN, HLS4ML, and Vitis AI.
This survey addresses these gaps by systematically mapping compression techniques to
FPGA-specific acceleration strategies and analyzing how compression choices directly impact
resource utilization (LUTs, DSPs, and BRAM) and performance metrics (throughput, latency, and energy
efficiency).

\subsection{Case Study Master Comparison}
\label{sec:mastertable}
Table~\ref{tab:master} normalizes all 25 case studies reviewed in this survey along a common
set of dimensions: reported technique category, target model/domain, hardware platform,
compression ratio, accuracy change, throughput, energy efficiency, DSP/LUT/BRAM usage, and 
baseline against which each result was compared. The \textbf{Cat.} column anticipates the
taxonomy developed in Section~\ref{sec:taxonomy}: \textbf{A} = DSP-Eliminating, \textbf{B} =
DSP-Repurposing/Mixed-Precision, \textbf{C} = Sparsity-Exploiting, \textbf{D} =
Memory-Hierarchy-Driven, \textbf{E} = Toolchain/Deployment-Level. Cells marked ``N/R'' indicate
a metric that the surveyed source does not report numerically (as opposed to reporting a
qualitative claim); ``N/A'' indicates the metric does not apply to that study's contribution
type (e.g., an analytical framework has no accuracy delta); ``--'' marks studies implemented as
an ASIC rather than an FPGA, included here because their compression-hardware co-design
techniques are directly comparable to the FPGA case studies. As the table is immediately
visible, only a minority of studies report a complete row: throughput, energy efficiency, and
DSP/LUT/BRAM utilization are the three most frequently missing fields, which is evidence
for the ``poorly characterized design trade-off'' gap discussed in
Section~\ref{sec:discussion}.

\begin{sidewaystable*}[p]
\centering
\caption{Case Study Master Comparison across the 25 works reviewed in this survey. ``N/R'' = not
reported numerically in the surveyed source; ``N/A'' = not applicable to this study's
contribution type; ``--'' = implemented as an ASIC, not an FPGA.}
\label{tab:master}
\resizebox{\textheight}{!}{%
\begin{tabular}{lccp{2.6cm}p{2.6cm}p{3.0cm}p{2.4cm}p{3.4cm}p{2.6cm}p{3.2cm}p{3.0cm}}
\toprule
\textbf{Work} & \textbf{Yr} & \textbf{Cat.} & \textbf{Domain} & \textbf{Platform} & \textbf{Compression Ratio} & \textbf{Acc.\ $\Delta$} & \textbf{Throughput} & \textbf{Energy Eff.} & \textbf{DSP/LUT/BRAM} & \textbf{Baseline Compared} \\
\midrule
Zhang et al.\ 2015~\cite{zhang2015optimizing} & 2015 & D & CNN (general) & N/R & N/A & N/A & N/R & N/R & N/R & N/A (analytical framework) \\
Venieris et al.\ 2016 -- fpgaConvNet~\cite{venieris2016fpgaconvnet,venieris2018fpgaconvnet} & 2016 & E & CNN (general) & N/R & N/A & N/R & N/R & N/R & N/R & N/A \\
Umuroglu et al.\ 2017 -- FINN~\cite{umuroglu2017finn} & 2017 & A & CNN (AlexNet/MNIST/CIFAR-10) & Xilinx ZC706 & $32\times$ (244\,MB$\to$7.4\,MB) & 95.8\% (MNIST) / 80.1\% (CIFAR-10) & 12.3M fps (MNIST) & 21\,W & 0 DSPs (XNOR-popcount) & Prior work ($48\times$ faster) \\
Blott et al.\ 2018 -- FINN-R~\cite{blott2018finn} & 2018 & B & CNN (general) & N/R & 2--16 bit (mixed precision) & N/R & N/R & N/R & Precision-tiered LUT/DSP mapping & N/A \\
Aimar et al.\ 2018 -- NullHop~\cite{aimar2018nullhop} & 2018 & C & CNN (general) & -- (28\,nm ASIC) & Act.\ sparsity 57\%$\to$82\% & N/R & 471.6\,GOp/s @ 500\,MHz & 155\,mW; 3\,TOp/s/W & -- (ASIC) & Dense baseline ($368\%$ eff.\ gain) \\
Han et al.\ 2017 -- ESE~\cite{han2017ese} & 2017 & C & LSTM (speech recognition) & N/R & $20\times$ ($10\times$ pruning $\times$ $2\times$ quant.) & N/R & N/R & N/R & N/R & N/A \\
Chen et al.\ 2016 -- Eyeriss~\cite{chen2016eyeriss} & 2016 & D & CNN (general) & -- (65\,nm ASIC) & N/A & N/A & N/R & Qualitative only (data movement $\gg$ compute cost) & -- (ASIC) & N/A \\
Li et al.\ 2020 -- FTRANS~\cite{li2020ftrans} & 2020 & C & Transformer & N/R & N/R (block-circulant; no ratio given) & N/R & N/R & N/R (qualitative gain) & N/R & CPU, GPU (qualitative) \\
Tian et al.\ 2025 -- On-FPGA Training~\cite{tian2025ultra} & 2025 & D & Transformer (training) & AMD Alveo U50 & N/R (low-rank tensor decomposition) & ``no accuracy loss'' (qualitative) & N/R & Qualitative gain vs.\ GPU & On-chip BRAM/URAM (no count given) & GPU training \\
Kabir et al.\ 2024 -- ProTEA~\cite{kabir2024protea} & 2024 & E & Transformer (BERT-style) & Xilinx Alveo & N/A (dense, uncompressed) & N/A & N/R & N/R & N/R & GPU + specialized FPGA baselines (qualitative) \\
Fahim et al.\ 2021 -- HLS4ML~\cite{fahim2021hls4ml} & 2021 & E & CNN/MLP (physics triggers) & FPGA + 65\,nm ASIC (energy figure) & QAP: 80\% sparsity, $50\times$ fewer bit ops & QAT @ 6-bit $\approx$ PTQ @ 14-bit & Sub-microsecond latency & 2.38\,nJ/inference (65\,nm ASIC) & 0 DSPs for binary layers & PTQ (14-bit) \\
Sharma et al.\ 2016 -- DNNWeaver~\cite{sharma2016dnnweaver} & 2016 & E & CNN (general) & Zynq, Stratix~V, Arria~10 & N/A & N/A & N/R & Qualitative (perf./W advantage) & Up to 100\% BRAM & CPUs, mid-range GPUs \\
Xilinx Vitis AI 2020~\cite{xilinx2020vitisai} & 2020 & E & CNN/general (TF, PyTorch, ONNX) & Xilinx DPU-based boards & N/R & N/R & N/R & N/R & N/R & N/A \\
Han et al.\ 2015 -- Deep Compression~\cite{han2015deep} & 2015 & A/C (hybrid) & CNN (AlexNet, VGG-16) & N/A (hardware-agnostic) & $35$--$49\times$ & ``no accuracy loss'' (qualitative) & N/R & $3$--$7\times$ (batch size 1) & N/A & Uncompressed baseline ($3$--$4\times$ speedup) \\
Ji et al.\ 2026 -- SpMM~\cite{ji2026spmm} & 2026 & C & LLM (chatglm2/LLaMA-2/Mixtral) & Intel Stratix~10~NX & $43$--$77\%$ effective sparsity & N/R & $3.89\times/2.85\times/2.66\times$ vs.\ dense GEMM & N/R & N/R & Dense GEMM; prior SpMM hardware \\
Liu et al.\ 2026 -- CXL-SpecKV~\cite{liu2026cxlspeckv} & 2026 & D & LLM (LLaMA-2, 7B--70B) & FPGA-attached CXL memory & $4$--$8\times$ KV capacity; $3$--$4\times$ KV compression & 99.5\% preservation & 1.6\,TB/s @ 800\,MHz; $3.2\times$ vs.\ GPU & $2.8\times$ mem.\ cost reduction & N/R & GPU-only baseline \\
Sun et al.\ 2026 -- HGQ~\cite{sun2026hgq} & 2026 & B & CNN/MLP (CERN L1 trigger) & N/R (via hls4ml/da4ml) & ``orders-of-magnitude'' (qualitative) & N/R & Sub-microsecond latency & N/R & Orders-of-magnitude reduction (qualitative) & Fixed-precision QAT \\
Hoang et al.\ 2026 -- KANEL\'{E}~\cite{hoang2026kanele} & 2026 & A & KAN (symbolic regr., HalfCheetah) & N/R & $>4000\times$ resource reduction & N/R & $>800$\,MHz clock; $2700\times$ speedup & N/R & 0 BRAM/DSP (LUT-only) & Prior KAN-on-FPGA \\
Cassidy et al.\ 2025 -- ReducedLUT~\cite{cassidy2025reducedlut} & 2025 & A & MLP (MNIST, jet substructure) & N/R & Up to 39\% P-LUT reduction & ``no measurable loss'' (qualitative) & N/R & N/R & Up to 39\% fewer P-LUTs & CompressedLUT (NeuraLUT) \\
Khataei \& Bazargan 2025 -- TreeLUT~\cite{khataei2025treelut} & 2025 & A & GBDT (MNIST, JSC, NID; tabular) & N/R & N/A (area-delay metric used instead) & ``comparable accuracy'' (qualitative) & N/R & N/R & LUT-only; $4$--$101\times$ lower area-delay product & PolyLUT-Add, NeuraLUT, DWN \\
Weng et al.\ 2025 -- AmigoLUT~\cite{weng2025amigolut} & 2025 & A & MLP ensembles (LUT-based) & N/R & $>10\times$ LUT reduction & ``equivalent accuracy'' (qualitative) & N/R & N/R & $>10\times$ fewer LUTs & Single large LogicNets/NeuraLUT model \\
Liu et al.\ 2025 -- FlightVGM~\cite{liu2025flightvgm} & 2025 & B & Diffusion Transformer (video gen.) & AMD V80 & N/R (DSP58: 2$\times$FP16 or 4$\times$INT8/DSP) & N/R & $1.30\times$ vs.\ RTX 3090 & $4.49\times$ vs.\ RTX 3090 & 2--4$\times$ MACs packed per DSP58 & NVIDIA RTX 3090 GPU \\
Zeng et al.\ 2024 -- FlightLLM~\cite{zeng2024flightllm} & 2024 & D & LLM (LLaMA2-7B) & Alveo U280; Versal VHK158 & Instr.\ storage 1.67\,TB$\to$3.25\,GB & N/R & HBM util.\ 35.6\%$\to$65.9\%; $>$A100 (VHK158) & $6.0\times$ vs.\ V100S & N/R & NVIDIA V100S; NVIDIA A100 \\
Gao et al.\ 2024 -- ESDA~\cite{gao2024esda} & 2024 & C & Event-based vision (DVS) & Xilinx ZCU102 & N/A (dynamic, data-dependent sparsity) & N/R & $54.8\times$ vs.\ embedded GPU; $10.2\times$ vs.\ NullHop & N/R & N/R & Embedded GPU; NullHop \\
Gerlinghoff et al.\ 2024 -- TLMAC~\cite{gerlinghoff2024tlmac} & 2024 & A & CNN (ResNet-18, ImageNet-scale) & Xilinx Virtex UltraScale+ & $3$--$12\times$ fewer LUTs & 71.8\% top-1 (full-precision accuracy retained) & N/R & N/R & LUT-only; $3$--$12\times$ fewer LUTs & LUTNet, Logic Shrinkage \\
\bottomrule
\end{tabular}%
}
\end{sidewaystable*}

\subsection{Taxonomy Overview}
The 25 case studies normalized in Table~\ref{tab:master} span the 2015--2026 publication range
and target CNNs, LSTMs, transformers, LLMs, Kolmogorov--Arnold Networks (KANs), and
gradient-boosted decision trees. Rather than presenting them chronologically, we group them by
their dominant \emph{hardware consequence}: the FPGA resource (DSP blocks, memory
hierarchy, or control/routing logic), and each work's compression-hardware co-design choice
primarily reshapes. Five categories emerged, identified by the letter used in Table~\ref{tab:master}
\textbf{Cat.} column. \textbf{(A) DSP-Eliminating} strategies entirely remove DSP blocks from the datapath, typically via binarized or LUT-tabulated arithmetic. \textbf{(B) The DSP-Repurposing and
Mixed-Precision} strategies retain DSP blocks but reconfigure how narrow-precision operations are
packed or time-multiplexed within them. \textbf{(C) Sparsity-Exploiting} strategies skip
computation on zero-valued weights or activations, whether statically pruned or dynamically
data-dependent. \textbf{(D) Memory-Hierarchy-Driven} strategies primarily target the on-/off-chip
data movement, not the arithmetic units themselves. \textbf{(E) Toolchain and
Deployment-Level} strategies are automated frameworks whose principal contribution is the
model-to-hardware mapping process itself. Several works legitimately span two categories: most
notably FlightLLM, which combines a DSP-repurposing sparse chain with a memory-hierarchy decode
scheme. In such cases we assign the category matching the mechanism most emphasized in the
surveyed source and cross-reference the secondary contribution where relevant.

\subsection{DSP-Eliminating Strategies}
\label{sec:taxA}
DSP-eliminating strategies remove multiply-accumulate DSP blocks from the datapath by reducing
arithmetic to Boolean logic (XNOR-popcount) or by tabulating an entire neuron, edge, or ensemble
member directly into a lookup table (LUT), so that FPGA logic fabric, not the DSP-limited
compute budget, becomes the governing resource constraint. FINN established the pattern with
binarized XNOR-popcount arithmetic; the remaining works below extend it to a physically-tabulated
LUT-neuron paradigm, for which the LUT count and routing congestion, not the DSP count, are the
determining efficiency metrics.

\subsubsection{Umuroglu et al.\ 2017 -- FINN~\cite{umuroglu2017finn}}
FINN demonstrated extreme compression--hardware co-design using 1-bit weights and
activations, achieving $32\times$ memory reduction (AlexNet: 244\,MB $\to$ 7.4\,MB) and fitting
entire models in an on-chip BRAM. A heterogeneous streaming architecture provides per-layer
custom hardware; XNOR-popcount arithmetic eliminates all DSP usage, and batch normalization
is folded into compile-time threshold comparisons (saving 2 DSPs, 55 FFs, 40 LUTs per
neuron). On Xilinx ZC706, FINN achieves 12.3 million fps on MNIST at 21\,W, 48$\times$ faster
than prior work, at the cost of accuracy: 95.8\% on MNIST versus 80.1\% on CIFAR-10.
Table~\ref{tab:finn} summarizes the key techniques used.

\begin{table}[!t]
\caption{FINN Framework: Key Techniques and Optimizations~\cite{umuroglu2017finn}}
\label{tab:finn}
\centering
\begin{tabular}{p{2cm}p{1.3cm}p{2.2cm}p{2cm}}
\toprule
\textbf{Technique} & \textbf{Type} & \textbf{Implementation} & \textbf{Key Benefit} \\
\midrule
1-bit quantization & Compression & Weights \& activations $\in \{-1,+1\}$ & $32\times$ memory reduction; fits on-chip BRAM \\
\midrule
XNOR-popcount & Hardware & Replace MAC with XNOR + popcount & Zero DSP usage; LUT savings \\
\midrule
BN folding & Both & Pre-compute BN as thresholds & Saves 2 DSPs, 55 FFs, 40 LUTs/neuron \\
\midrule
Boolean pooling & Hardware & Max-pool as OR gates & Minimal logic; no arithmetic \\
\midrule
Streaming arch. & Hardware & Dedicated PE per layer; on-chip streams & Eliminates DRAM; 36,400 ops/byte \\
\midrule
Folding & Hardware & Time-multiplex using $P$ PEs, $S$ SIMD lanes & Scalable: 12K--12.3M fps \\
\bottomrule
\end{tabular}
\end{table}

\subsubsection{Gerlinghoff et al.\ 2024 -- TLMAC~\cite{gerlinghoff2024tlmac}}
TLMAC directly extends the LUT-based computing-in-memory family surveyed
above (LogicNets, PolyLUT, NeuraLUT, KANEL\'{E}) from small, fully unrolled
networks to ImageNet-scale models by exploiting weight redundancy instead of the 
per-neuron LUT absorption. Because a purely bit-parallel LUT mapping of a
multiply-accumulate scale as $2^{G \cdot B_a}$ in the number of grouped
activation inputs, TLMAC instead serializes computation bit-by-bit across
input bits, reducing the LUT count to grow only linearly with the weight bit width
while repurposing unused LUT-6 input lines as a selection signal that lets a
single small LUT array store multiple quantized weight groups from a
2--4 bit network and switch between them at runtime. The place-and-route stage
first uses spectral clustering to losslessly compress the sequential weight
dimension into the minimum number of LUT arrays, and then applies simulated
annealing is applied to reduce the inter-LUT routing congestion by up to 50\%. Applied to
Nonuniform-to-Uniform-Quantised ResNet-18 at 2--4 bits on a Xilinx Virtex
UltraScale+ FPGA, TLMAC fits an entire ImageNet-scale model in soft logic with
full-precision accuracy (71.8\% top-1) using 3--12$\times$ fewer LUTs than
prior binary-network LUT approaches (LUTNet, Logic Shrinkage). This demonstrates
that the toolchain fragmentation and manual-optimization gaps identified for LUT-based DNNs (Section~\ref{sec:discussion}) can be
substantially closed by treating weight-group redundancy, not neuron
absorption, as the primary compression lever.

\subsubsection{Cassidy et al.\ 2025 -- ReducedLUT~\cite{cassidy2025reducedlut}}
ReducedLUT targets the LUT-based neural network family introduced by LogicNets and
NeuraLUT, in which an entire neuron or sub-network is tabulated directly into a physical
lookup table (P-LUT). Because the resulting truth tables are trained on data instead of
being designed by hand, large regions of the input space are never observed during training;
ReducedLUT marks these unobserved entries as ``don't cares'' and reassigns their values
to maximize self-similarity among the sub-tables produced by the CompressedLUT
decomposition, thereby reducing the number of unique sub-tables that must be
physically implemented. An \emph{exiguity} parameter bounds the number of dependent
sub-tables that must be re-derived when a given entry is modified, trading the search
runtime for compression. On NeuraLUT benchmarks (MNIST, jet substructure
classification), ReducedLUT reduces P-LUT utilization by up to 39\% relative to
CompressedLUT with no measurable accuracy loss, demonstrating that unstructured
``don't-care'' exploitation is a complementary axis of compression to the
quantization and pruning techniques discussed in Section~\ref{sec:background}: one
that acts purely on the synthesized hardware representation, not on the
trained model itself.

\subsubsection{Khataei and Bazargan 2025 -- TreeLUT~\cite{khataei2025treelut}}
TreeLUTs depart from the DNN-centric focus of most literature surveyed here,
instead mapping gradient boosted decision trees (GBDTs) to FPGAs as a
competitive alternative for tabular classification tasks. Because a decision
tree is structurally a binary decision diagram, it maps naturally onto the LUT
fabric without the exponential fan-in blowup that constrains LogicNets-style
LUT-based DNNs. TreeLUT quantizes thresholds by training XGBoost directly on
pre-quantized features (avoiding costly quantization-aware training) and
quantizes leaf values using a per-tree bias-and-scale reformulation that
minimizes bitwidth, then compiles the ensemble into a fully unrolled
three-layer (key generator, decision trees, adder tree) pipelined Verilog
architecture using only LUTs, with no BRAMs or DSPs. Across MNIST, JSC, and
network-intrusion-detection benchmarks, TreeLUT achieves a 4--101$\times$ lower
area-delay product than recent LUT-based NNs (PolyLUT-Add, NeuraLUT, DWN) at
comparable accuracy, illustrating that for tabular data the model class itself,
not just its compression, determines hardware efficiency: a consideration
largely absent from the compression-technique taxonomy in
Section~\ref{sec:background}.

\subsubsection{Weng et al.\ 2025 -- AmigoLUT~\cite{weng2025amigolut}}
AmigoLUT addresses the scalability limitation shared by all LUT-based DNN
accelerators surveyed above (LogicNets, PolyLUT, NeuraLUT, KANEL\'{E}): LUT usage
grows exponentially with neuron fan-in, thus increasing accuracy by widening or
deepening a single network, which quickly exhausts FPGA resources. Instead of scaling
a single model, AmigoLUT ensembles many small LUT-based networks, showing
empirically (via a novel diversity-plot analysis and the disagreement-error-ratio
metric) that simple averaging outperforms bagging and AdaBoost for heavily
quantized, sparse models. The key hardware challenge is combining ensemble
members whose learned input/output quantization scales differ; AmigoLUT resolves
this with a hybrid scheme: a single shared low-precision input quantizer
followed by per-member re-quantization, and an output-side layer that aligns all
members to a common scale before summation. This avoids the per-member
multiply-accumulate rescaling required by naive ensembling. This yields
an order-of-magnitude reduction in LUT usage compared to a single large
LogicNets/NeuraLUT model at equivalent accuracy, showing that ensembling is a
viable, linearly-scaling alternative to the exponential cost of widening
individual LUT-based architectures.

\subsubsection{Hoang et al.\ 2026 -- KANEL\'{E}~\cite{hoang2026kanele}}
KANEL\'{E} introduced the first systematic FPGA deployment framework for
Kolmogorov--Arnold Networks (KANs), a class of neural networks that replaces the
fixed activation functions and matrix multiplications of MLPs with learnable
spline-based edge functions, inspired by the Kolmogorov--Arnold representation theorem.
This activation-centric structure is inherently suited to LUT-based FPGA implementation:
each spline, defined over a fixed quantized domain, can be independently pruned and mapped
to a single LUT, reducing network inference to LUT lookups and additions with zero BRAM
or DSP usage. Quantization-aware training and pruning are co-optimized during the training
phase; the additive independence of the KAN edges makes unstructured pruning hardware-friendly
in a way that is fundamentally incompatible with conventional LUT-based DNNs. Implemented
across multiple benchmarks including symbolic regression, scientific computing, and
continuous control (HalfCheetah), KANEL\'{E} operates above 800\,MHz, achieves a
state-of-the-art area--delay product, and delivers up to 2700$\times$ speedup and over
4000$\times$ resource reduction compared with prior KAN-on-FPGA implementations, establishing
that architecture-hardware co-design, a theme central to this survey, can be extended to
novel network paradigms beyond standard DNNs. Most recently, LUT-LLM extended the same
DSP-eliminating, LUT-based paradigm from small unrolled networks to billion-parameter LLMs,
replacing linear-layer arithmetic with table lookups over pre-computed dot products on an
AMD~V80 FPGA; however,the scale of application the LUT-based family surveyed above has not previously been
reached~\cite{he2026lutllm}.

\subsection{DSP-Repurposing and Mixed-Precision Strategies}
\label{sec:taxB}
DSP-repurposing and mixed-precision strategies retain DSP blocks but change how narrow-precision
operations are packed within them, addressing the mismatch between a DSP slice's native
fixed-width multiplier (e.g., $27\times18$-bit) and a network's much narrower learned bit-width.
Bit Fusion~\cite{sharma2018bit} (Section~\ref{sec:background}) established the core
mechanism, packing multiple narrow multiplications into one wide DSP slice, which was extended to automated per-layer bit-width search (FINN-R), DSP58 hybrid-precision packing
for diffusion transformers (FlightVGM), and fully learned per-parameter bit-width assignment
(HGQ). Two related mixed-precision approaches fall outside our 25 normalized case studies but
are worth noting here: FILM-QNN assigns precision at filter granularity within a single layer,
combining a majority of low-precision filters with a minority of high-precision filters to recover
accuracy lost to aggressive quantization~\cite{sun2022filmqnn}, and FINN-GL extends FINN's
mixed-precision datapath from feedforward CNNs to recurrent LSTM cells via an ONNX
Scan-operator formulation, addressing the recurrent-network gap in FINN's original
scope~\cite{khandelwal2025finngl}.

\subsubsection{Blott et al.\ 2018 -- FINN-R~\cite{blott2018finn}}
FINN-R extends FINN to arbitrary-precision quantization (2--16 bits) with automated design
space exploration. A three-layer architecture (frontend, IR, backend) enables mixed-precision
optimization, assigning 2--4 bit quantization to insensitive layers and 8-bit to critical layers.
Precision-specific arithmetic maps 2--4 bit layers to 16-entry LUT multiplication, 4--8 bit to
LUT-carry chains, and 12--16 bit to DSP blocks, with folding factors scaled inversely with
precision to maintain a balanced pipeline throughput.

\subsubsection{Liu et al.\ 2025 -- FlightVGM~\cite{liu2025flightvgm}}
FlightVGM extends the compression-hardware co-design theme of this survey from
CNNs and LLMs to Diffusion Transformer (DiT)-based video generation models
(VGMs), a compute-bound workload with over 40$\times$ the operational intensity
of LLM inference. It combines three techniques directly analogous to those
discussed in Section~\ref{sec:background}: (1) an online \emph{activation
sparsification} architecture that exploits spatial and temporal token
similarity across video frames, functioning as a runtime, data-dependent analog
of the structured pruning discussed earlier; (2) a \emph{hybrid-precision DSP58
expansion} that packs either two FP16 or four INT8 MACs per DSP via bit
overpacking, addressing the same fixed-precision DSP limitation that motivated
FINN-R's mixed-precision datapath; and (3) a dynamic-static scheduler that
reorders operators at runtime to absorb the load imbalance introduced by
sparsification, echoing the load-balancing challenge identified for ESE in
Section~\ref{sec:taxonomy}. Implemented on an AMD V80 FPGA, FlightVGM
surpasses an NVIDIA 3090 GPU by 1.30$\times$ in throughput and 4.49$\times$ in
energy efficiency despite a $>$21$\times$ peak-compute disadvantage, reinforcing
this survey's central claim that co-designed sparsity and precision can close a
large raw-compute gap against GPUs.

\subsubsection{Sun et al.\ 2026 -- HGQ~\cite{sun2026hgq}}
HGQ (High Granularity Quantization) extends quantization-aware training beyond the
per-layer or per-channel granularity of frameworks such as QKeras and FINN-R, assigning
an independent learnable bit-width to each individual weight and activation via gradient
descent. A differentiable fixed-point quantization scheme relaxes inherently discrete
bit-widths to continuous surrogates during training, using the Straight-Through Estimator
to propagate gradients; parameters assigned zero bit-width are automatically pruned,
unifying quantization and unstructured sparsity within a single training objective. A
jointly trained differentiable on-chip resource estimator penalizes LUT- and DSP-heavy
configurations as a regularizer, explicitly trading accuracy against FPGA resource usage
without manual sensitivity analysis, and addressing the mixed-precision optimization burden
we identify as an open challenge. Integrating with hls4ml and da4ml for hardware
synthesis, HGQ delivers orders-of-magnitude reduction in resource consumption and latency
compared to conventional fixed-precision QAT, enabling sub-microsecond inference for
applications including CERN ATLAS and CMS Level-1 trigger particle physics systems,
where hard resource caps and strict latency constraints make automated per-parameter
co-optimization essential.

\subsection{Sparsity-Exploiting Strategies}
\label{sec:taxC}
Sparsity-exploiting strategies skip computation on zero-valued weights or activations instead
of reducing arithmetic precision. A central tension recurs across this category:
\emph{unstructured} sparsity (Deep Compression's magnitude-based pruning, ESE's per-weight
pruning) achieves higher compression ratios but produces irregular non-zero distributions that
cause load imbalance on regular hardware, whereas \emph{structured} approaches (FTRANS's
block-circulant representation, Ji et al.'s block-aggregated attention sparsity) sacrifice some
compression ratio to preserve the regular, GEMM-compatible access patterns that FPGA PE arrays
require. NullHop and ESDA further distinguish \emph{static} weight sparsity, fixed at compile
time, from \emph{dynamic, data-dependent} activation sparsity,  which varies at runtime and must be
handled with online zero-skipping logic. A related building block outside our normalized case
studies is Systolic Sparse Tensor Slices, which augment the systolic array itself with 2D
in-fabric blocks supporting multiple degrees of sparsity, allowing a single array to serve both
sparse and dense workloads without requiring a separate sparse datapath~\cite{taka2025systolic}.

\subsubsection{Han et al.\ 2015 -- Deep Compression~\cite{han2015deep}}
Deep Compression achieves 35--49$\times$ compression on AlexNet and VGG-16 without
accuracy loss via a three-stage pipeline: (1) magnitude-based pruning ($9$--$13\times$);
(2) $k$-means weight sharing with 8-bit indices for convolutional layers; and (3) Huffman
coding for an additional 20--30\% reduction. AlexNet shrinks from 240\,MB to 6.9\,MB; and
VGG-16 from 552\,MB to 11.3\,MB. At batch size 1, compressed networks achieve 3--4$\times$
speedup and 3--7$\times$ energy efficiency improvement.

\subsubsection{Han et al.\ 2017 -- ESE~\cite{han2017ese}}
ESE, the first sparse LSTM accelerator on FPGA, achieves $20\times$ compression via
load-balance-aware pruning ($10\times$) and 12-bit quantization ($2\times$). Unlike standard
pruning which creates unbalanced non-zero distributions and causes PE idle time, load-balance-aware
pruning enforces equal sparsity quotas across submatrices, improving utilization from 60--80\%
to $>$90\% at 90\% sparsity. Per-PE FIFO activation queues decouple fast PEs from slow ones,
maintaining high utilization despite the irregular sparsity.

\subsubsection{Aimar et al.\ 2018 -- NullHop~\cite{aimar2018nullhop}}
NullHop exploits activation sparsity through zero-skipping and, operates on compressed sparse
feature maps to avoid redundant MAC operations. Quantization increased the average activation
sparsity from 57\% (FP32) to 82\% (16-bit fixed-point), enabling up to a 368\% computational
efficiency improvement. A zero-skipping pipeline maintains $>$98\% MAC utilization whereas a
binary sparsity map stores only nonzero values. Implemented in 28\,nm at 16-bit,
NullHop achieves 471.6\,GOp/s at 500\,MHz consuming 155\,mW (3\,TOp/s/W).

\subsubsection{Li et al.\ 2020 -- FTRANS~\cite{li2020ftrans}}
FTRANS accelerates transformer inference on FPGAs by replacing dense weight matrices with
block-circulant representations, thereby enabling FFT-based matrix multiplication where circulant
operations are reduced to element-wise frequency-domain multiplications. Dedicated FFT engines
and pipelined executions exploit this structure, achieving significant performance and energy
improvements over CPU and GPU baselines for transformer inference.

\subsubsection{Gao et al.\ 2024 -- ESDA~\cite{gao2024esda}}
The ESDA targets event-based vision, where DVS cameras produce spatially sparse
frame data whose sparsity ratio and pattern vary dynamically at runtime, unlike
the static weight sparsity emphasized elsewhere in this survey. The ESDA comprises
an all-on-chip dataflow accelerator from parametrizable modules that share a
unified sparse token-feature interface, streaming only non-zero coordinate-value
pairs through cascaded convolution modules instead of dense feature maps. At
the algorithm level, submanifold sparse convolution constrains the output non-zero
locations to match the input, preventing the ``dilation'' effect of standard
convolution that would otherwise densify activations layer by layer, as
demonstrated in Section~\ref{sec:taxonomy}'s discussion of NullHop and ESE:
exploiting sparsity in hardware only pays off if the algorithm preserves it.
A sparsity-aware hardware/algorithm co-optimization flow estimates per-module
latency and BRAM/DSP cost directly from dataset-level sparsity statistics,
formulating the load-balanced resource allocation problem as mixed-integer
programming in the same spirit as Zhang et al.'s roofline analysis. On an
embedded ZCU102 FPGA, ESDA achieved up to 54.8$\times$ speedup over an embedded
GPU and 10.2$\times$ over NullHop, illustrating that dynamic, data-dependent
sparsity, not just static compression, can be a first-class target for
FPGA dataflow co-design.

\subsubsection{Ji et al.\ 2026 -- SpMM for Sparse Attention~\cite{ji2026spmm}}
Ji et al.\ addressed the challenge of running sparse attention efficiently on GEMM-optimized
FPGA hardware, where the dynamic data routing required by conventional SpMM accelerators
is incompatible with static, high-frequency PE arrays of devices such as the Intel
Stratix~10~NX with hardened Tensor Blocks. Their key observation is that non-zero attention
values form rectangular clusters because of the structural properties of the attention mechanism,
enabling a block aggregation strategy that dynamically merges unpruned values into compact
dense GEMM-like tiles. An index merge-sort module transforms block-pruned sparse attention
matrices into regular tiles that fit directly onto PE arrays using the same static data flow
as dense GEMM, eliminating complex routing logic and preserving the clock frequency and
scalability. Exploiting 43--77\% effective sparsity in the attention-value product across
popular LLMs, the design achieves throughput gains of 3.89$\times$, 2.85$\times$, and
2.66$\times$ over a dense GEMM baseline on chatglm2-6b-32k, LLaMA-2-7b-chat-4k, and
Mixtral-8$\times$7b, respectively, and up to 2.68$\times$ higher throughput versus prior
SpMM hardware, demonstrating that structured sparsity exploitation, a central challenge
identified in our discussion, can be made compatible with GEMM-optimized FPGA datapaths
through appropriate algorithmic co-design.

\subsection{Memory-Hierarchy-Driven Strategies}
\label{sec:taxD}
Memory-hierarchy-driven strategies treat data movement, not arithmetic throughput, as the
primary bottleneck, following directly from Zhang et al.'s roofline observation that many FPGA
CNN layers are memory-bandwidth-limited, not compute-limited. Eyeriss's Row Stationary
dataflow and Tian et al.'s fully on-chip training pipeline both minimize off-chip transfers
through data reuse and hierarchy-aware scheduling, whereas FlightLLM and CXL-SpecKV extend the
same principle to the datacenter LLM setting, where the ``memory wall'' takes the form of
decode-stage activation bandwidth and KV-cache capacity instead of CNN weight reuse.

\subsubsection{Zhang et al.\ 2015 -- Roofline Model~\cite{zhang2015optimizing}}
Zhang et al.\ introduced a roofline-based analytical framework for optimizing FPGA CNN
accelerators, revealing that many layers are memory-bandwidth limited and, not compute
limited. The framework applies loop tiling, loop unrolling/pipelining, and loop transformations
to balance the computational throughput against the memory bandwidth. This robust analysis directly
motivated subsequent aggressive compression work by demonstrating that reducing the model memory
footprint, not just increasing computation, is key to achieving peak FPGA performance.

\subsubsection{Chen et al.\ 2016 -- Eyeriss~\cite{chen2016eyeriss}}
Eyeriss minimizes energy by reducing data movement through the Row Stationary dataflow:
weights are reused across images, inputs are shared across output channels, and partial sums are kept
local until complete, across four memory levels (off-chip DRAM, shared buffer, network-on-chip,
and PE-local storage). The key finding is that data movement costs far more energy than
arithmetic, making intelligent dataflows more important than raw compute throughput.

\subsubsection{Zeng et al.\ 2024 -- FlightLLM~\cite{zeng2024flightllm}}
The FlightLLM is the direct predecessor of the FlightVGM discussed above and tackles the
same DSP-precision mismatch problem for text-only transformer LLMs. It
introduces a configurable sparse DSP chain (CSD-Chain) that divides a long
cascaded DSP48 chain into groups linked by a reconfigurable path with a sparse
multiplexer, a reduction node, and an overflow-adjustment unit, allowing the same
hardware to compute dense GEMM/GEMV or arbitrary N:M sparse matrix operations
without the $\sim$5$\times$ area penalty of prior sparse-DSP designs. To address
the decode-stage memory bottleneck, FlightLLM proposes an always-on-chip decode
scheme that fuses computation across an entire inference step so activations
never leave on-chip buffers between layers, raising HBM bandwidth utilization
from 35.6\% to 65.9\%, directly extending the memory-hierarchy optimization
theme established by DNNBuilder and Ma et al.\ in Section~\ref{sec:background}
to the transformer-decode setting. A length-adaptive compilation scheme further
reuses instructions across neighboring token lengths, shrinking instruction
storage from an infeasible $\sim$1.67\,TB to 3.25\,GB. On a Xilinx Alveo U280,
FlightLLM achieves 6.0$\times$ higher energy efficiency and 1.8$\times$ better
cost efficiency than an NVIDIA V100S GPU on LLaMA2-7B, and its Versal
VHK158 implementation surpasses an NVIDIA A100 in throughput, evidence that
the FPGA-specific co-design principles surveyed here scale from CNN/LSTM
accelerators to billion-parameter generative language models.

\subsubsection{Tian et al.\ 2025 -- On-FPGA Transformer Training~\cite{tian2025ultra}}
Tian et al.\ developed the first FPGA accelerator for end-to-end transformer training using 
bi-directional contraction flow that decomposes weight matrices into compact low-rank tensors.
All parameters, activations, and gradients fit within the on-chip BRAM and UltraRAM, eliminating the
off-chip transfers. when tested on an AMD Alveo U50, the system achieved major gains in memory
efficiency and energy over GPU training without accuracy loss.

\subsubsection{Liu et al.\ 2026 -- CXL-SpecKV~\cite{liu2026cxlspeckv}}
CXL-SpecKV directly targets the memory capacity bottleneck that our survey identifies as a
persistent challenge for large transformer deployment: the KV-cache of an LLaMA-2 70B
model serving 2048 tokens at batch size~32 consumes 640\,GB, far exceeding the GPU
capacity. Liu and Yu propose a disaggregated KV-cache architecture combining three
innovations: (1)~a CXL-based memory offloading framework that expands the effective KV-cache
capacity 4--8$\times$ by storing cache entries in FPGA-attached memory connected via the
CXL~2.0 cache-coherent protocol; (2)~a lightweight LSTM-based speculative prefetcher that
predicts and preloads future token cache entries with 95\% accuracy, hiding CXL access
latency; and (3)~an FPGA cache engine operating at 800\,MHz that implements 3--4$\times$
KV-cache compression and decompression at 1.6\,TB/s throughput without consuming GPU
computing resources. when evaluated on 7B--70B LLMs integrated with vLLM and TensorRT-LLM,
CXL-SpecKV achieves 3.2$\times$ higher throughput, 2.8$\times$ memory cost reduction,
and 99.5\% accuracy preservation over GPU-only baselines, illustrating that FPGA-based
memory management can complement model compression to address the memory wall in
datacenter inference. SpeedLLM applies the same memory-hierarchy-driven principle to
Alveo~U280-hosted LLM inference outside our normalized case studies, combining data-stream
parallelism, an operator-fusion scheme for LLaMA2, and a memory-reuse strategy to reduce
off-chip traffic~\cite{wang2025speedllm}.

\subsection{Toolchain and Deployment-Level Strategies}
\label{sec:taxE}
Toolchain and deployment-level strategies are distinguished from the previous four categories by
their primary contribution being the \emph{mapping process} from a trained model to
synthesizable hardware, rather than a single algorithmic or hardware mechanism. FINN
(Section~\ref{sec:taxA}) is arguably also a toolchain in this sense, but is categorized under
DSP-Eliminating above because its binarization mechanism is a more distinctive contribution
relative to the works below, which instead differentiate themselves primarily by automation
philosophy: fpgaConvNet and DNNWeaver auto-generate custom per-model hardware from a dataflow or
macro-ISA specification; Vitis AI deploys to a fixed, pre-synthesized DPU overlay; HLS4ML
spatially unrolls an entire (typically small) network for sub-microsecond latency, and ProTEA
targets runtime-programmable dense transformers without per-model hardware regeneration.
Table~\ref{tab:toolchain} and Section~\ref{sec:discussion} return to this category for a deeper
automation-vs-performance comparison.

\subsubsection{Venieris et al.\ 2016 -- fpgaConvNet~\cite{venieris2016fpgaconvnet,venieris2018fpgaconvnet}}
fpgaConvNet pioneered the streaming dataflow architecture where each layer has dedicated
hardware that passes data through on-chip streams, eliminating external memory bottlenecks. Its
Synchronous Dataflow-based framework explores designs through graph partitioning, coarse-grained
folding, and fine-grained folding. Graph partitioning enables large networks on
resource-constrained FPGAs but incurs $\sim$10-second reconfiguration latency, limiting its
applicability to batch processing workloads.

\subsubsection{Sharma et al.\ 2016 -- DNNWeaver~\cite{sharma2016dnnweaver}}
DNNWeaver generates synthesizable FPGA accelerators from high-level DNN specifications in the
Caffe format via a macro dataflow ISA. Its Template Resource Optimization algorithm
co-optimizes the accelerator configuration and execution schedule by partitioning layer computations
into on-chip-fitting slices, mapping PE buffers to BRAMs, and mapping arithmetic units to DSP slices.
Evaluated across Xilinx Zynq, Altera Stratix V, and Arria 10, it delivers substantial
performance-per-watt advantages over CPUs and mid-range GPUs with up to 100\% BRAM
utilization.

\subsubsection{Xilinx Vitis AI 2020~\cite{xilinx2020vitisai}}
Vitis AI provides an integrated toolchain for the quantization and compilation of pre-synthesized
Deep Learning Processing Units (DPUs), supporting TensorFlow, PyTorch, and ONNX. The
Vitis AI Quantizer performs PTQ or QAT on INT8, whereas the compiler applies operator fusion
and workload partitioning. Because DPU overlays are pre-built for target FPGAs, model updates
require only recompiling the instruction file instead of hardware regeneration and, trading peak
performance for rapid, practitioner-accessible deployment.

\subsubsection{Fahim et al.\ 2021 -- HLS4ML~\cite{fahim2021hls4ml}}
HLS4ML is an open-source co-design workflow that bridges ML frameworks with FPGA/ASIC hardware
via HLS. Integrating with QKeras, it supports arbitrary-precision fixed-point arithmetic (binary
to 16-bit); QAT maintains accuracy at 6-bit whereas PTQ requires 14-bit. Quantization-Aware
Pruning (QAP) achieved 80\% sparsity on 6-bit models with $50\times$ fewer bit operations.
The XNOR-Popcount eliminates DSP usage for binary layers, and compile-time BN folding reduces
runtime overhead, yielding sub-microsecond latency and 2.38\,nJ/inference on 65\,nm ASIC.

\subsubsection{Kabir et al.\ 2024 -- ProTEA~\cite{kabir2024protea}}
The ProTEA targets standard dense transformers with runtime programmability, allowing users to
specify configurations (number of heads and, layers) at runtime without hardware rebuilding.
Architecture-aware matrix tiling with distinct tile sizes for attention and feedforward blocks
keeps the computing units busy. Evaluated on Xilinx Alveo with BERT-style models, ProTEA
outperformed both GPU baselines and specialized FPGA accelerators.

\section{Quantitative Meta-Analysis}
\label{sec:metaanalysis}
Table~\ref{tab:master} normalizes the 25 case studies along a common set of column
\emph{headings}; however, a heading shared across rows does not imply that the underlying numbers are
directly comparable. This section makes explicit which quantities can and cannot be placed on
a common axis, and presents the resulting comparisons, deliberately sparse where the
literature itself is sparse, instead of papering over missing or incommensurable data.

\subsection{Normalization Methodology}
\label{sec:normalization}
Four obstacles prevented a single unified comparison across all the 25 case studies.
\textbf{Throughput units differ irreconcilably}: works report fps (FINN, ESDA), GOp/s
(NullHop), TOp/s (KANEL\'{E}'s clock rate only), GFLOPS, or relative speedup factors against
an unspecified baseline clock, and conversion between them would require per-model operation
counts that are not stated in the surveyed sources. \textbf{Energy-efficiency units differ by
what is held constant}: some works report an absolute rate (NullHop's 3\,TOp/s/W; HLS4ML's
2.38\,nJ/inference), while others report only a multiplier against a named baseline device
(FlightVGM's $4.49\times$ vs.\ an RTX 3090; FlightLLM's $6.0\times$ vs.\ a V100S); the two
families cannot be placed on one axis without assuming a specific baseline device's absolute
efficiency, which this survey does not have independent grounds to supply.
The \textbf{``compression ratio'' was measured against at least three different baselines}
depending on the work: (i) an uncompressed FP32 model (Deep Compression's $35$--$49\times$;
FINN's $32\times$), (ii) a prior \emph{compressed} competitor instead of an uncompressed model
(ReducedLUT vs.\ CompressedLUT; AmigoLUT and TLMAC vs.\ prior LUT-binary methods), or (iii) a
non-weight quantity entirely, such as the KV-cache size (CXL-SpecKV) or instruction storage
(FlightLLM). A ratio from family (ii) plotted against a ratio from family (i) visually
implies a comparison that the underlying numbers do not support.
\textbf{Accuracy is reported as either an absolute value or a delta, inconsistently}: FINN
reports absolute accuracy achieved (95.8\% MNIST / 80.1\% CIFAR-10) without stating the FP32
baseline accuracy it is compressing from, so no loss/delta can be derived for it; TLMAC,
ReducedLUT, and AmigoLUT instead report a qualitative delta directly (``retained,'' ``no
measurable loss,'' ``equivalent accuracy'') without the absolute number. Given these four
obstacles, Sections~\ref{sec:ivb} and~\ref{sec:ivc} plot only the subset of studies for which a
single, stated baseline underlies both axes of the same figure, and label each plotted point
with that baseline explicitly instead of assuming comparability. This incommensurability
is a field-wide symptom, not an artifact of our particular 25-study sample,
corroborated by Kratos, a standardized FPGA benchmark suite built specifically to allow unrolled
DNNs to be compared across sparsity levels and precisions on a common basis, an effort motivated
by the same lack of shared baselines in this section ~\cite{dai2024kratos}.

\subsection{Compression Ratio vs.\ Accuracy Change}
\label{sec:ivb}
\begin{figure}[tbp]
\centering
\begin{tikzpicture}
\begin{axis}[
  width=0.95\columnwidth,
  height=5.2cm,
  xlabel={Compression ratio (vs.\ study's own stated baseline, $\times$)},
  ylabel={Accuracy change (\%, negative = loss)},
  xmin=0, xmax=55,
  ymin=-3, ymax=3,
  grid=both,
  grid style={gray!20},
  legend pos=south east,
  legend style={font=\scriptsize},
  tick label style={font=\scriptsize},
  label style={font=\scriptsize}
]
\addplot[only marks, mark=square*, mark size=3pt, color=blue,
  error bars/.cd, x dir=both, x explicit]
  coordinates {(42,0) +- (7,0)};
\addlegendentry{Deep Compression (weights, vs.\ FP32)}
\addplot[only marks, mark=triangle*, mark size=3.5pt, color=red,
  error bars/.cd, x dir=both, x explicit]
  coordinates {(3.5,-0.5) +- (0.5,0)};
\addlegendentry{CXL-SpecKV (KV-cache, vs.\ uncompressed cache)}
\end{axis}
\end{tikzpicture}
\caption{Compression ratio vs.\ accuracy change for the only two of 25 case studies reporting
both quantities against a single stated baseline. Horizontal bars show each study's reported
range. Note the two points compress fundamentally different quantities (model weights vs.\
KV-cache) and are not on a common scale; they are plotted together only because both happen to
satisfy the single-baseline criterion of Section~\ref{sec:normalization}.}
\label{fig:compaccuracy}
\end{figure}
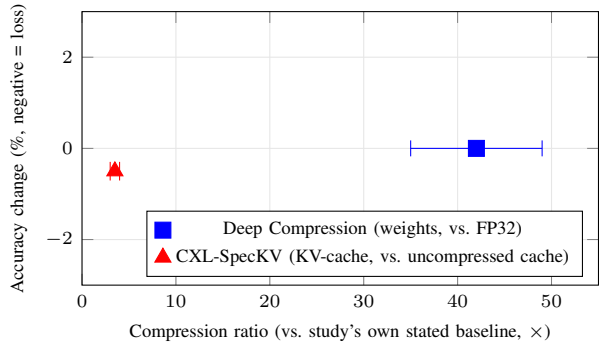

Of the 25 case studies in Table~\ref{tab:master}, only two report a compression ratio and an
accuracy figure measured against a single, stated baseline within the same study: \textbf{Deep
Compression}~\cite{han2015deep}, whose $35$--$49\times$ parameter-compression ratio (pruning +
weight-sharing + Huffman coding, vs.\ the original FP32 AlexNet/VGG-16) is reported alongside
``no accuracy loss''; and \textbf{CXL-SpecKV}~\cite{liu2026cxlspeckv}, whose $3$--$4\times$
KV-cache compression ratio (vs.\ an uncompressed KV-cache) is reported alongside 99.5\%
accuracy preservation (a $0.5$\% loss). These two points do \emph{not} measure the same kind
of compression: one compresses the model weights and, the other compresses inference-time
key-value cache. Figure~\ref{fig:compaccuracy} shows this distinction explicitly with
different marker shapes instead of treating them as a single trend line. Four further 
studies (TLMAC, ReducedLUT, AmigoLUT, KANEL\'{E}) all report near-zero accuracy change but are
excluded from the figure because their compression ratios are measured against a
\emph{different prior compressed method}, not a common uncompressed baseline, per
Section~\ref{sec:normalization}; all four independently reported near-zero loss while using
incompatible reference frames is itself a finding, discussed further below. The remaining 19
studies report neither quantity in a form comparable to that of a common baseline.

\subsection{Energy Efficiency Across Categories}
\label{sec:ivc}
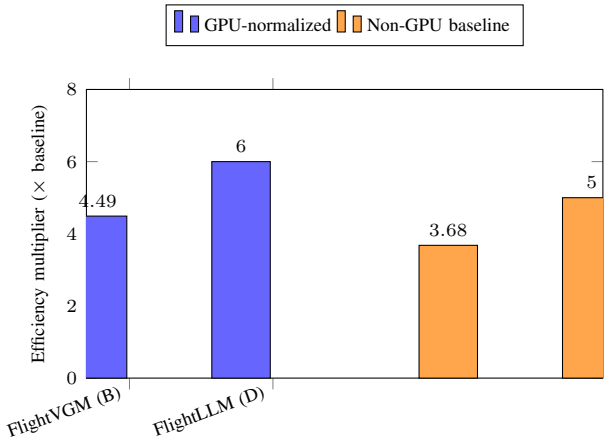
\begin{figure}[tbp]
\centering
\begin{tikzpicture}
\begin{axis}[
  width=0.95\columnwidth,
  height=5.4cm,
  ybar,
  bar width=22pt,
  symbolic x coords={FlightVGM (B), FlightLLM (D), NullHop (C), Deep Compr.\ (C)},
  xtick=data,
  x tick label style={font=\scriptsize, rotate=20, anchor=east},
  ylabel={Efficiency multiplier ($\times$ baseline)},
  label style={font=\scriptsize},
  tick label style={font=\scriptsize},
  ymin=0, ymax=8,
  nodes near coords,
  nodes near coords style={font=\scriptsize},
  every node near coord/.append style={/pgf/number format/.cd, fixed, precision=2},
  legend style={font=\scriptsize, at={(0.5,1.15)}, anchor=south, legend columns=2}
]
\addplot[fill=blue!60, draw=black] coordinates {(FlightVGM (B),4.49) (FlightLLM (D),6.0)};
\addplot[fill=orange!70, draw=black] coordinates {(NullHop (C),3.68) (Deep Compr.\ (C),5.0)};
\legend{GPU-normalized, Non-GPU baseline}
\end{axis}
\end{tikzpicture}
\caption{Reported energy-efficiency multipliers for the four (of 25) case studies that provide
one, grouped by baseline type instead of forced into a per-category distribution that the data
cannot support. Deep Compression's bar uses the midpoint of its reported $3$--$7\times$ range.
Category letters in parentheses refer to Table~\ref{tab:master}.}
\label{fig:energyeff}
\end{figure}

Only two case studies have reported energy efficiency as a multiplier against a named GPU baseline:
\textbf{FlightVGM}~\cite{liu2025flightvgm} (Category B, DSP-Repurposing), at $4.49\times$ versus
an NVIDIA RTX~3090, and \textbf{FlightLLM}~\cite{zeng2024flightllm} (Category D,
Memory-Hierarchy-Driven), at $6.0\times$ versus an NVIDIA V100S. Two further studies reported an
efficiency multiplier against a \emph{non-GPU} baseline: \textbf{NullHop}~\cite{aimar2018nullhop}
(Category C; an ASIC, not an FPGA) reports up to $3.68\times$ (368\%) against a dense,
non-sparse baseline on the same silicon, and \textbf{Deep Compression}~\cite{han2015deep}
(Category C) reports $3$--$7\times$ against its own uncompressed baseline at a batch size of~1.
Figure~\ref{fig:energyeff} plots all four, visually distinguishing GPU-normalized bars from
non-GPU-normalized bars; no bar chart ``grouped by category'' in the sense of showing a
distribution per category is possible, since at most two studies populate any single category
with a directly comparable number. The remaining 21 studies reported energy efficiency only
qualitatively (e.g., ``substantial performance-per-watt advantage,'' DNNWeaver) or not at all.

\subsection{Where the Literature Disagrees}
\label{sec:disagree}
Two genuine tensions emerge from the case studies reviewed in Section~\ref{sec:taxonomy}:
this survey cannot fully reconcile with the information available in the surveyed sources.

First, Koilia and Kachris~\cite{koilia2024hardware} reported that FPGA throughput advantages over
GPUs for LLM inference \emph{narrow as sequence length grows}, a workload-dependent finding
discussed in Section~\ref{sec:discussion}. FlightLLM~\cite{zeng2024flightllm}. In contrast, it has been
reported that its Versal VHK158 implementation \emph{surpasses} an NVIDIA A100 in throughput on
LLaMA2-7B. These are not necessarily contradictory; FlightLLM's always-on-chip decode
scheme is a specific architectural response that could plausibly hold its advantage at longer
sequence lengths than a generic FPGA baseline would. However, the surveyed FlightLLM source does
not report throughput as a function of sequence length, so whether its result falls inside or
outside the regime where Koilia and Kachris observe FPGA advantage narrowing cannot be
determined from the text available to this survey. We call this unresolved and, not
reconciled.

Second, ESE~\cite{han2017ese} and NullHop~\cite{aimar2018nullhop} demonstrated that
load-balance-aware pruning and zero-skipping, respectively, can recover high hardware
utilization ($>$90\% for ESE) from sparse computation \emph{using dedicated scheduling
hardware}. In contrast, ProTEA~\cite{kabir2024protea}, demonstrates that a runtime-programmable
\emph{dense} accelerator, avoiding sparsity-induced control overhead entirely, can match
or exceed specialized sparse designs while supporting a broader range of models. This is a
direct disagreement about strategy, not simply a difference in reported numbers: ESE/NullHop's
implicit position is that sparsity is worth exploiting if the scheduling problem is solved,
whereas ProTEA's implicit position is that avoiding the scheduling problem altogether is
preferable. Because ESE and NullHop target LSTM/CNN workloads while ProTEA targets dense
BERT-style transformers, the survey lacks a single apples-to-apples comparison that would
settle which strategy dominates for a shared workload. We flag this as an open question
for future benchmarking (Table~\ref{tab:agenda}) instead of resolving it here.

\section{Discussion}
\label{sec:discussion}

\subsection{Compression Techniques: A Comparative Perspective}
Quantization offers the best balance between compression and hardware friendliness: INT8 achieves
$4\times$ compression with $<$0.5\% accuracy loss~\cite{jacob2018quantization} and maps
cleanly to DSP slices, making it the preferred method for production deployment. Pruning
typically yields 5--10$\times$ compression with 1--2\% accuracy loss~\cite{blalock2020state}; however, unstructured sparsity requires specialized scheduling hardware to avoid PE underutilization.
Structured pruning preserves the architectural regularity at the cost of lower compression ratios,
making it better suited for systolic and dataflow FPGA implementations. Binary
networks~\cite{umuroglu2017finn} achieve $32\times$ compression at the cost of a 3--4\%
accuracy drop, whereas hybrid pipelines combining pruning, quantization, and Huffman coding
reach 35--49$\times$ with $<$1\% loss~\cite{han2015deep}. Knowledge distillation produces
2--10$\times$ smaller models retaining 95--98\% teacher accuracy~\cite{gou2021knowledge},
and can be combined with quantization to further reduce hardware cost without significant
accuracy degradation. Low-rank factorization delivers 5--7$\times$ compression but is less
widely adopted because it alters the layer structure, complicates hardware mapping, and often
fails to yield a real speedup without custom kernel support~\cite{yi2021towards}. Overall, INT8
quantization dominates at moderate compression levels; hybrid strategies that combine multiple
techniques are the most effective when compression ratios beyond $10\times$ are required. These
trends are consistent with the broader compression literature: chronological taxonomies of the
field~\cite{liu2025survey,dantas2024comprehensive} likewise identify quantization as the most
hardware-friendly technique and hybrid pipelines as necessary beyond moderate compression
ratios, while domain-specific surveys of NLP~\cite{gupta2022compression}, large language
models~\cite{zhu2024survey}, and transformer architectures~\cite{tang2024survey} report the
same accuracy--cost escalation pattern for their respective model families, suggesting that
trade-offs observed here for FPGA-targeted CNNs and transformers generalize across domains,
although these surveys do not analyze hardware-specific implications.
Table~\ref{tab:technique_summary} presents the ratio/accuracy/hardware-fit figures cited above
from a single side-by-side view.

\begin{table}[!t]
\centering
\caption{Compression Techniques: Ratio, Accuracy Cost, and FPGA Hardware Fit}
\label{tab:technique_summary}
\resizebox{\columnwidth}{!}{%
\begin{tabular}{lp{1.5cm}p{1.6cm}p{3.2cm}}
\toprule
\textbf{Technique} & \textbf{Compression Ratio} & \textbf{Accuracy Cost} & \textbf{FPGA Hardware Fit} \\
\midrule
INT8 quantization~\cite{jacob2018quantization} & $4\times$ & $<$0.5\% & Maps cleanly to DSP slices; preferred for production \\
\midrule
Structured pruning~\cite{blalock2020state} & 5--10$\times$ & 1--2\% & Preserves regularity; suited to systolic/dataflow designs, lower ratio than unstructured \\
\midrule
Unstructured pruning~\cite{blalock2020state} & 5--10$\times$ & 1--2\% & Higher potential density but needs specialized scheduling hardware to avoid PE underutilization \\
\midrule
Binarization~\cite{umuroglu2017finn} & $32\times$ & 3--4\% & Eliminates DSPs entirely via XNOR-popcount; LUT-based parallelism \\
\midrule
Hybrid (prune+quant+Huffman)~\cite{han2015deep} & 35--49$\times$ & $<$1\% & Best ratio/accuracy trade-off; needed once ratios exceed $10\times$ \\
\midrule
Knowledge distillation~\cite{gou2021knowledge} & 2--10$\times$ & 2--5\% (95--98\% retained) & Combines with quantization; no direct DSP/LUT mapping of its own \\
\midrule
Low-rank factorization~\cite{yi2021towards} & 5--7$\times$ & N/R & Alters layer structure; often needs custom kernels for real speedup \\
\bottomrule
\end{tabular}%
}
\end{table}

\subsection{Hardware Implications of Compression Choices}
Compression choices cascade directly into FPGA resource utilization, memory hierarchy design,
and datapath complexity. Binarization eliminates DSP blocks entirely via XNOR-popcount,
redirecting resources to LUT-based parallelism~\cite{umuroglu2017finn}, INT8 maps efficiently
to DSP slices with $4\times$ memory reduction~\cite{xilinx2020vitisai}, and mixed-precision
assigns 2--4 bit layers to 16-entry LUT multipliers, 4--8 bit to LUT-carry chains, and
12--16 bit to DSP blocks at reduced parallelism~\cite{blott2018finn}. On the memory side,
aggressive compression such as FINN's $32\times$ reduction enables complete on-chip BRAM
storage and streaming at 36,400\,ops/byte, fully eliminating DRAM bottlenecks. Instead, moderate
compression demands careful tiling, data reuse, and partitioning
strategies~\cite{sharma2016dnnweaver,zhang2015optimizing} to stay within the on-chip memory
limits. Eyeriss~\cite{chen2016eyeriss} demonstrated that even with compression, data movement
costs far more energy than arithmetic, making intelligent dataflow design a first-order concern
alongside the compression ratio. Unstructured sparsity requires load-balance enforcement across
processing elements~\cite{han2017ese} and zero-skipping pipelines with binary sparsity
maps~\cite{aimar2018nullhop} to translate theoretical compression into real hardware efficiency
gains, and PE idle time can reduce utilization from over 90\% to
60--80\%.

\subsection{Acceleration Strategy Trade-offs}
The choice of the acceleration architecture must be co-designed with the compression strategy.
Systolic arrays provide high computational density for dense, regular workloads, but degrade
under irregular sparsity patterns where zero-skipping disrupts rigid data flow synchronization.
Streaming dataflow designs~\cite{umuroglu2017finn,venieris2016fpgaconvnet} offer
heterogeneous per-layer hardware specialization and on-chip pipelining, achieving high
throughput with zero DRAM traffic; however, they consume substantial BRAM for inter-layer buffering
and may underutilize resources in narrow layers. Sparse accelerators such as
ESE~\cite{han2017ese} and NullHop~\cite{aimar2018nullhop} realize efficiency only when
sparsity is structured or load-balanced; otherwise, the control overhead for scheduling and
irregular memory access negates theoretical gains. Runtime-programmable dense accelerators,
such as ProTEA~\cite{kabir2024protea},  demonstrate that avoiding irregular control can entirely
match or exceed specialized sparse designs while supporting diverse model architectures. Static
FPGA reconfiguration yields optimal per-segment throughput but incurs $\sim$10-second
latency~\cite{venieris2016fpgaconvnet}, restricting its use to batch workloads, and dynamic
time-multiplexing in FINN-R~\cite{blott2018finn} trades throughput for deployment flexibility
without reconfiguration overhead. These trade-offs confirm that compression strategy selection
should account not only for model size reduction but also for the downstream hardware
complexity it introduces. This pattern is consistent with cross-platform hardware surveys:
Deng et al.~\cite{deng2020model} similarly find that no single accelerator architecture
dominates across CPU, GPU, FPGA, and ASIC targets, and Koilia and Kachris~\cite{koilia2024hardware}
report that FPGA throughput advantages over GPUs for LLM inference narrow as sequence length
grows, reinforcing that accelerator choice must be workload-specific, not fixed. Silvano
et al.~\cite{silvano2025survey} extended this observation to heterogeneous HPC platforms, where
the same dataflow-versus-density trade-offs identified above for individual FPGAs recur at
the cluster scale.

\subsection{Deployment Toolchains and Workflows}
\label{sec:toolchains}
The four AI-to-FPGA toolchains examined in Section~\ref{sec:taxE} span the full
automation--performance spectrum, as summarized quantitatively in Table~\ref{tab:toolchain}.
\textbf{FINN}~\cite{umuroglu2017finn,blott2018finn} generates fully customized streaming
hardware per model, achieving the highest throughput via XNOR-popcount datapaths, on-chip
weight storage, and per-layer parallelism tuning; however, it requires hours of HLS synthesis and
place-and-route, with hardware regeneration required for any model or quantization change.
\textbf{HLS4ML}~\cite{fahim2021hls4ml} spatially unrolls the entire network across the FPGA
fabric, allocating dedicated hardware per neuron to achieves sub-microsecond latency; it scales
only to $\sim$100K parameters, making it ideal for latency-critical scientific and edge control
applications and not for large networks. \textbf{Vitis AI}~\cite{xilinx2020vitisai} deploys
models to pre-synthesized DPU overlays in minutes, supporting INT8 across TensorFlow,
PyTorch, and ONNX without hardware regeneration, at the cost of lower peak throughput
compared to custom datapaths. \textbf{DNNWeaver}~\cite{sharma2016dnnweaver} sits between
FINN and Vitis AI on the automation spectrum: its Template Resource Optimization algorithm
automates accelerator configuration and scheduling from a Caffe specification; however, like
FINN, it still produces a custom per-model design instead of targeting a fixed overlay, so a
new model requires a new synthesis instead of a simple recompilation step. A critical shared
limitation is that all four frameworks treat training, compression, and hardware mapping as
sequential, not co-optimized, steps, leaving significant joint optimization
opportunities unexplored. Manual intervention also remains necessary across all four:
selecting QAT versus PTQ, tuning per-layer bit-widths and reuse factors, and restructuring
models to avoid operators that require CPU fallback. A unified intermediate representation
carrying compression metadata and hardware constraints is  a significant step toward
closing this gap.

\begin{table}[!t]
\centering
\caption{Toolchain Comparison. ``N/R'' = not reported/quantified for this toolchain in the
surveyed source.}
\label{tab:toolchain}
\resizebox{\columnwidth}{!}{%
\begin{tabular}{lp{2.1cm}p{2.1cm}p{2.1cm}p{2.3cm}}
\toprule
& \textbf{FINN} & \textbf{HLS4ML} & \textbf{Vitis AI} & \textbf{DNNWeaver} \\
\midrule
\textbf{Automation level} & Low: custom per-model synthesis, manual folding/buffer tuning & High for small nets: fully automated spatial unrolling & High: fixed pre-synthesized DPU overlay, no regeneration & Automated accelerator generation (Template Resource Optimization), but custom per-model synthesis \\
\midrule
\textbf{Target precision range} & 1-bit (binarized) & Binary--16-bit (arbitrary fixed-point, via QKeras) & INT8 (PTQ or QAT) & N/R \\
\midrule
\textbf{Reconfiguration cost} & Hours (HLS synthesis + place-and-route per model/quantization change) & N/R (HLS-based; cost not quantified in surveyed source) & Minutes (recompile to DPU instructions only) & N/R (custom synthesis per model; cost not quantified) \\
\midrule
\textbf{Peak throughput regime} & Highest of the four: 12.3M fps on MNIST @ 21\,W (XNOR-popcount, on-chip weights) & Sub-microsecond latency; 2.38\,nJ/inference (65\,nm ASIC) & Lower peak throughput than custom datapaths (qualitative) & Substantial perf./W advantage over CPUs/mid-range GPUs (qualitative); up to 100\% BRAM utilization \\
\midrule
\textbf{Model scale limit} & N/R (not explicitly bounded; demonstrated on AlexNet-scale compressed models) & $\sim$100K parameters (explicit) & N/R (not explicitly bounded; supports arbitrary TF/PyTorch/ONNX models via DPU) & N/R \\
\midrule
\textbf{Framework support} & N/R (binarized-network-specific; input framework not named in surveyed source) & QKeras & TensorFlow, PyTorch, ONNX & Caffe \\
\bottomrule
\end{tabular}%
}
\end{table}

Two additional Category~E toolchains discussed in Section~\ref{sec:taxE}, fpgaConvNet and
ProTEA, are omitted from Table~\ref{tab:toolchain} because they represent automation
philosophies orthogonal to the precision/regeneration axes compared above: fpgaConvNet's
defining cost is its $\sim$10-second inter-partition reconfiguration
latency~\cite{venieris2016fpgaconvnet}, not a precision range, and ProTEA's defining
property is runtime configurability of a dense transformer \emph{without any hardware
regeneration at all}~\cite{kabir2024protea}, which does not fit a ``reconfiguration cost'' cell
in the same units as the other four.

\subsection{Related Deployment Tools Beyond This Survey's Scope}
Beyond FINN, HLS4ML, and Vitis AI, several other toolchains pursue the same goal of training a
network on FPGA hardware. On the industrial side, \textbf{Intel's OpenVINO toolkit}~\cite{intel2024openvino} provides an
FPGA/VPU inference runtime conceptually parallel to Vitis AI's DPU-based flow, targeting
Intel FPGA platforms with a similar train-once, deploy-via-overlay philosophy.
\textbf{Xilinx DNNDK}~\cite{xilinx2019dnndk} was the direct predecessor to Vitis AI, offering
the same fixed-DPU deployment model prior to Vitis AI's broader framework and quantizer
support. On the compiler side, \textbf{TVM with the VTA backend}~\cite{chen2018tvm,moreau2019vta}
offers a framework-agnostic compiler stack that can lower models to a configurable tensor
accelerator on FPGA, occupying a middle ground between FINN's fully custom per-model hardware
and Vitis AI's fixed overlay by allowing the accelerator template itself to be resynthesized.
\textbf{HeteroCL}~\cite{lai2019heterocl} and \textbf{PyLog}~\cite{huang2021pylog}
are general-purpose, Python-based HLS abstraction layers, not ML-specific pipelines,
which play a role analogous to hls4ml's high-level description-to-HLS bridge, but target
arbitrary reconfigurable-computing workloads instead of being specialized for neural network
deployment. Two further tools sit closer to DNNWeaver's fully automated, per-model hardware
generation philosophy than to the overlay-based approaches above. \textbf{AutoSA}~\cite{wang2021autosa}
is a polyhedral-compilation-based tool that automatically generates systolic array
accelerators for FPGA from a high-level algorithmic specification by, applying space-time
transformations to map computation onto a configurable systolic template. It occupies a
similar niche to TVM/VTA's resynthesizable tensor accelerator but derives the mapping through
polyhedral analysis instead of compiler auto-tuning, which targets general linear-algebra and
tensor kernels instead of being restricted to a fixed operator set. \textbf{DeepBurning}~\cite{wang2016deepburning}
predates DNNWeaver as an early automated framework for generating FPGA-based learning
accelerators, assembling a target architecture from a library of pre-designed RTL hardware
building blocks for common neural-network operators and generating the corresponding
compiler for a given network topology, and automating per-model hardware generation in the same
spirit as DNNWeaver's Template Resource Optimization algorithm, but via building-block
assembly instead of dataflow-ISA scheduling. A systematic head-to-head evaluation of these
tools alongside FINN, HLS4ML, and Vitis AI is left for future research.

\section{Open Challenges and Research Agenda}
\label{sec:agenda}
Despite substantial progress, several critical gaps remain in the literature. Table~\ref{tab:agenda} formalizes
each as a three-part entry: the current state-of-the-art attempt at the problem (as reviewed in
Section~\ref{sec:taxonomy}), why that attempt falls short of fully resolving it, and a concrete
next step grounded in extending an existing, cited technique, not a purely speculative
direction.

\begin{table*}[!t]
\centering
\caption{Research Agenda: Open Challenges, Current State-of-the-Art Attempts, and Concrete Next Steps}
\label{tab:agenda}
\resizebox{\textwidth}{!}{%
\begin{tabular}{p{2.6cm}p{4.8cm}p{5.0cm}p{5.2cm}}
\toprule
\textbf{Challenge} & \textbf{Current SOTA Attempt} & \textbf{Why Insufficient} & \textbf{Concrete Next Step} \\
\midrule
Toolchain fragmentation &
FINN-R's frontend/IR/backend architecture supporting multiple ML frameworks~\cite{blott2018finn}; Vitis AI's native TensorFlow/PyTorch/ONNX support~\cite{xilinx2020vitisai} &
Each toolchain still requires manual format conversion and framework-specific quantization conventions when moving between FINN, HLS4ML, and Vitis AI (Table~\ref{tab:toolchain}); no shared representation carries compression metadata across tools &
A unified, ONNX-style intermediate representation carrying per-layer bit-width, sparsity pattern, and target-FPGA resource constraints, so one compressed-model description can be lowered by any of the four toolchains without manual reconversion \\
\midrule
Accuracy--efficiency characterization &
FINN's own reported task-dependent spread: 95.8\% (MNIST) vs.\ 80.1\% (CIFAR-10) under identical binarization~\cite{umuroglu2017finn} &
This is a single toolchain's two-dataset data point, not a systematic study; none of the 25 works in Table~\ref{tab:master} map compression ratio to accuracy loss as a function of task complexity or dataset characteristics &
A shared benchmark suite evaluating the same compression technique across a graded difficulty ladder of tasks/datasets, reporting accuracy $\Delta$ as a function of task complexity rather than a single point estimate \\
\midrule
Automated mixed-precision optimization &
FINN-R's per-layer sensitivity analysis assigning 2--4 bit to insensitive layers, 8-bit to critical ones~\cite{blott2018finn}; HGQ's differentiable, per-parameter learned bit-width~\cite{sun2026hgq} &
FINN-R still depends on manual/heuristic per-layer sensitivity analysis; HGQ automates the per-parameter case but is demonstrated only on hls4ml/da4ml-integrated, physics-trigger-scale models, not shown as a general replacement for FINN-R-style analysis on large CNNs/transformers &
Extend HGQ's differentiable, resource-aware bit-width learning beyond its current hls4ml/da4ml integration to the streaming architectures used by FINN and DNNWeaver \\
\midrule
Sparse computation reliability &
ESE's load-balance-aware pruning enforcing equal sparsity quotas across submatrices~\cite{han2017ese}; NullHop's zero-skipping pipeline with binary sparsity maps~\cite{aimar2018nullhop} &
Both are architecture-specific point solutions (LSTM; CNN, respectively); without their bespoke load-balancing, utilization drops from $>$90\% to 60--80\% (ESE) or requires dynamic control logic with irregular memory access (NullHop), and neither generalizes automatically to new sparsity patterns &
Generalize ESDA's mixed-integer-programming resource-allocation co-optimization~\cite{gao2024esda}, currently scoped to event-based vision, into a hardware/algorithm framework that jointly selects pruning granularity and PE scheduling from a target FPGA's DSP/BRAM budget \\
\midrule
Persistent memory bottlenecks &
CXL-SpecKV's disaggregated KV-cache via CXL-attached memory~\cite{liu2026cxlspeckv}; FlightLLM's always-on-chip decode scheme~\cite{zeng2024flightllm} &
Both target only the LLM decode/KV-cache bottleneck; CXL-SpecKV's gains partly depend on its speculative prefetcher holding its reported 95\% accuracy, and FlightLLM's scheme, demonstrated on LLaMA2-7B, does not resolve the roofline-identified~\cite{zhang2015optimizing} bandwidth bottleneck for training-time activation/gradient memory (cf.\ next row) &
Extend the on-chip/CXL memory-hierarchy techniques validated for LLM inference decode to the training setting, and characterize prefetcher/cache-hit sensitivity across a wider range of model sizes than currently reported \\
\midrule
FPGA-based training &
Tian et al.'s bi-directional contraction flow keeping all parameters, activations, and gradients on-chip for transformer training~\cite{tian2025ultra} &
Demonstrated only on a single AMD Alveo U50 at single-batch scale, using full-precision formats; FPGA training in general remains rare because its high memory-bandwidth and dynamic-computation-graph requirements conflict with FPGAs' static, inference-optimized architectures &
Hybrid CPU/FPGA training pipelines partitioning gradient computation (FPGA for compressed forward/backward passes, host for optimizer state), combined with quantization-aware training (Section~\ref{sec:background}) adapted for stable low-precision gradient updates rather than inference only \\
\bottomrule
\end{tabular}%
}
\end{table*}

Five broader research themes recur across this agenda (hardware-aware neural architecture
search, adaptive per-input precision, architecture-specific compression, on-device training
support, and standardized benchmarking), not as independent items but as threads running
through Table~\ref{tab:agenda}: hardware-aware NAS and adaptive
precision both bear on the mixed-precision and sparsity rows, architecture-specific
compression bears on the memory-bottleneck row (e.g.\ FTRANS's and Tian et al.'s structured
representations~\cite{li2020ftrans,tian2025ultra}), on-device training support \emph{is} the
FPGA-based-training row, and standardized benchmarking is the connective tissue the
accuracy--efficiency and toolchain-fragmentation rows call for.

\section{Conclusion}
\label{sec:conclusion}

This survey organized 25 FPGA compression-hardware co-design case studies (2015--2026) into a
five-category taxonomy defined by which FPGA resource each strategy primarily reshapes:
DSP-eliminating, DSP-repurposing/mixed-precision, sparsity-exploiting, memory-hierarchy-driven,
and toolchain/deployment-level, normalized along a common set of dimensions
(Table~\ref{tab:master}). This  normalization surfaced a field-wide characterization gap: only one
of the 25 works reported a compression ratio and accuracy change against a single common
baseline, and only two reported energy efficiency normalized against a common GPU baseline, so the
literature is characterized well enough to motivate individual designs but not yet well enough to
support the cross-study comparisons of a practitioner choosing between FINN, HLS4ML, Vitis AI, and
DNNWeaver. Building on this taxonomy and meta-analysis, we formalized six open
challenges (Table~\ref{tab:agenda}): toolchain fragmentation, accuracy--efficiency
characterization, automated mixed-precision optimization, sparse computation reliability,
persistent memory bottlenecks, and FPGA-based training, each paired with a concrete next step
grounded in extending an existing technique, which is not a general call for future work. As new
accelerators are published, we expect most to slot into one of these five categories, and we offer
the taxonomy and research agenda as a starting point for the common-baseline, cross-toolchain
comparisons this field does not yet have.

\section{Acknowledgment}
\label{sec:Acknowledgment}

Claude was used to correct the grammatical errors.

\end{document}